\documentclass[
preprint,
amsmath,
amssymb,
aps,
]{revtex4-1}

\usepackage{subfigure}
\usepackage{graphicx}
\usepackage{dcolumn}
\usepackage{bm}
\usepackage{color}
\usepackage{pifont}
\usepackage{hyperref} 

\begin{document}

\preprint{ITP307/2023-03}

\title{Minimum Connected Dominating Set and Backbone of a Random Graph}

\author{Yusupjan Habibulla$^{1,2}$}
\email{Email: yusupjan@alumni.itp.ac.cn; zhouhj@itp.ac.cn}

\author{Hai-Jun Zhou$^{3,4,5}$}
\email{Email: zhouhj@itp.ac.cn}

\affiliation{$^1$School of Physics and Technology, Xingjiang University,Huarui Street 777,Urumqi 830017, China}

\affiliation{$^2$Xinjiang Key Laboratory of Solid State Physics and Devices, Xinjiang University, Urumqi, 830017, China}

\affiliation{$^3$CAS Key Laboratory for Theoretical Physics, Institute of Theoretical Physics, Chinese Academy of Sciences, Beijing 100190, China}

\affiliation{$^4$MinJiang Collaborative Center for Theoretical Physics, MinJiang University, Fuzhou 350108, China}

\affiliation{$^5$School of Physical Sciences, University of Chinese Academy of Sciences, Beijing 100049, China}

\date{\today}

\begin{abstract}
We study the minimum dominating set problem as a representative combinatorial optimization challenge with a global topological constraint.  The requirement that the backbone induced by the vertices of a dominating set should be a connected subgraph makes the problem rather nontrivial to investigate by statistical physics methods. Here we convert this global connectivity constraint into a set of local vertex constraints and build a spin glass model with only five coarse-grained vertex states. We derive a set of coarse-grained belief-propagation equations and obtain theoretical predictions on the relative sizes of minimum dominating sets for regular random and Erd\"os-R\'enyi random graph ensembles. We also implement an efficient message-passing algorithm to construct close-to-minimum connected dominating sets and backbone subgraphs for single random graph instances. Our theoretical strategy may also be inspiring for some other global topological constraints.
\end{abstract}


\maketitle

\tableofcontents

\section{Introduction}

During the last two decades a lot of efforts have been made in applying the cavity method of spin glass physics to random combinatorial optimization problems. These interdisciplinary research efforts have greatly advanced our understanding on the complexity of these hard problems, and have inspired the invention of several message-passing strategies to tackle them~\cite{Mezard-etal-2002,Krzakala-etal-PNAS-2007,Mezard-Montanari-2009}. Most of these theoretical and algorithmic studies have focused on the effects of local constraints, such as the $K$-body constraints of the random $K$-satisfiability problem and the edge constraints of the random vertex-cover problem. Global constraints on the overall topological properties of the graph, such as requiring that all the vertices in a particular state should form a connected subgraph, are much more difficult to handle within the framework of cavity method. Some progress has been made in recent years along the research direction of building easy-to-handle spin glass models to tackle hard graphical problems with global constraints, such as converting the forest-subgraph constraint of the minimum feedback vertex set problem into a set of local edge constraints~\cite{Zhou-2013} and converting the $K$-core collapse constraint of the optimal $K$-core attack problem into a set of single vertex constraints~\cite{Zhou-2022,Zhou-Zhou-2023}. These local-interaction models have facilitated the design of efficient message-passing algorithms which only need to pass low-dimensional vector messages along the edges of the graph to guiding the search for optimal globally constrained microscopic configurations. An example of successful interdisciplinary applications was approximately solving the optimal dismantling problem of a complex network with roughly linear time complexity~\cite{Mugisha-Zhou-2016}.

Our present work is a continuation of this research line. We aim at building local constraints to realize the global connectivity constraint, and we take the connected dominating set (CDS) problem as a representative example system~\cite{Du-Wan-2013}.  If a subset $\Gamma$ of vertices of a graph has the property that every vertex of this graph is either a member of $\Gamma$ or is adjacent to a member of $\Gamma$, then it is referred to as a dominating set. A trivial example of a dominating set is the set which contains all the vertices of the graph, but often it is desirable to reduce the size of the dominating set in practical applications such as power-grid surveillance~\cite{Yang-Wang-Motter-2012}. If the subgraph $\mathcal{B}$ induced by all the vertices of a dominating set $\Gamma$ and all the edges between these vertices is a connected (i.e., there is a path of consecutive edges within this subgraph linking any two vertices of this subgraph), then the subset $\Gamma$ is a connected dominating set. We call the subgraph $\mathcal{B}$ that is uniquely determined from this connected dominating set a backbone of the graph. Notice that if the whole graph $G$ under study is not connected itself, it is impossible to construct a CDS for it and there is then no backbone for the whole graph. In this situation we could consider each connected component of the graph $G$ and discuss the issue of CDS and backbone separately for each such component. In the present work we assume the graph $G$ under study is indeed connected. A backbone $\mathcal{B}$ for such a connected graph $G$ then has a very nice property: Between any two vertices $i$ and $j$ of $G$ there exists at least one path with at most two edges not insider the backbone $\mathcal{B}$.

The optimization objective of the CDS problem is to construct a connected dominating set $\Gamma$ of size as small as possible, namely the minimum CDS problem. This is a problem belongs to the class NP-hard (nondeterministic polynomial hard)~\cite{GrayMRJohnsonDS1979,GuhaKhuller1998}. The minimum CDS problem has wide practical applications and is widely studied in computer science and operations research~\cite{Du-Wan-2013,SwamyCKumarA2004}. For example, in wireless ad-hoc networks, a connected dominating set can create a virtual network backbone for packet routing and control~\cite{2001Flooding,2002Dominating,2002Span}. The requirement that the vertices in the dominating set should form a connected subgraph (a graph backbone) is a global connectivity constraint involving all the vertices in the dominating set. This global constraint greatly increases the difficulty of tackling this problem by statistical physics methods. On the theoretical side, there is still no theoretical attempt to predict the minimum size of connected dominating sets for a given random graph ensemble. On the algorithmic side, the highly efficient message-passing algorithm of Ref.~\cite{Zhao-Habibulla-Zhou-2015} only works for the local constraints, and no message-passing algorithm has attempted to take into account this global connectivity constraint. 

In the present work we study the CDS problem by statistical physics methods and message-passing algorithms. First, we design a set of local vertex constraints to the global connectivity constraint and derive a set of coarse-grained belief-propagation equations. Our model for the CDS problem is then essentially a five-state Potts spin glass problem with local constraints. We then use this model to estimate the relative size of minimum CDS solutions for regular random graphs and Erd\"os-R\'enyi random graphs, and implement a {\tt BBQ} message-passing algorithm to construct approximately minimum CDS solutions for single random graph instances. The algorithmic results are close to the predicted minimum sizes. Our spin glass model on single graph instances could also be used by other optimization methods such as simulated annealing.  The idea of this present work is likely applicable to any general combinatorial optimization problems with a global connectivity constraint.

\section{The minimum backbone problem}
\label{sec:problemsetting}

Consider a graph (network) $G$ formed by $N$ vertices (nodes) and $M$ undirected edges (links) between these vertices. The vertices are indexed by positive integers $i, j, k, \ldots \in \{1, \ldots, N\}$. Two vertices $i$ and $j$ are said to be nearest neighbors (adjacent) if there is an edge between them, and this edge is denoted as $(i, j)$. The set of nearest neighbors of vertex $i$ is denoted as $\partial i$, namely $\partial i \equiv \{ j : (i, j) \in G\}$. The size of this set, $d_i \equiv |\partial i |$, is defined as the degree of vertex $i$. Without loss of generality we assume $G$ to be a connected graph so that there is a connected path of consecutive edges between any two vertices. (If $G$ is composed of two or more mutually disconnected components, each of these connected components shall be considered separately.)

A dominating set of graph $G$ is a subset of the $N$ vertices which satisfies the property that every vertex of $G$ not in this subset is adjacent to at least one vertex of this subset~\cite{2002Dominating,Yang-Wang-Motter-2012}. The concept of dominating set is closely related to the control strategies of complex networked systems~\cite{Liu-Barabasi-2016}. If the subgraph induced by a dominating set $\Gamma$ of a connected graph $G$, which contains all the vertices of $\Gamma$ and all the edges between these vertices, is itself connected, then $\Gamma$ is said to be a connected dominating set, and the induced subgraph is said to be a backbone of graph $G$. In other words, a backbone of a connected graph $G$ is a connected subgraph with the following adjacency property: every vertex $i$ of $G$ either belongs to this subgraph or is adjacent to at least one vertex of this subgraph.

Since there is a one-to-one correspondence between a backbone and a connected dominating set $\Gamma$, we will refer to a backbone by its (connected dominating) vertex set $\Gamma$ in later discussions.  With respect to a backbone $\Gamma$, we say a vertex $i$ is occupied if it belongs to this backbone and is unoccupied (or empty) if it does not belong to this backbone.

The connectedness constraint for a backbone $\Gamma$ ensures that information or materials could be transmitted from one vertex of $\Gamma$ to another vertex of $\Gamma$ through the internal edges of the backbone~\cite{Yeung-Saad-Wong-2013}. Since the backbone $\Gamma$ is adjacent to all the other vertices of graph $G$, it could serve as a highway of transmission between any two vertices~\cite{Shao-Zhou-2007}.  Let us denote by $E$ the number of vertices in a backbone. A minimum backbone is a backbone which contains the minimum number of vertices (i.e., $E$ achieves the minimum value),  and its vertex set $\Gamma$ is a minimum connected dominating set. For example, the minimum backbone for a star-shaped graph contains only one single vertex ($E= 1$), but that of a chain-shaped graph has size $E = (N-2)$. Exactly solving the minimum backbone problem for a general connected graph, which often contains an abundant number of loops, is a very difficult combinatorial optimization task. 

The present work aims at estimating the relative size $\rho = E / N$ of minimum backbones for different random graph ensembles by the mean field theory of statistical physics. We are also interested in designing efficient message-passing algorithms to construct close-to-minimum backbone solutions for single random graph instances. The requirement of the backbone being a connected subgraph is a global constraint, which makes this problem much more difficult to tackle than the conventional minimum dominating set problem~\cite{Zhao-Habibulla-Zhou-2015,Habibulla-Zhao-Zhou-2015,Habibulla-2020}. Concerning computational efficiency, a main technical challenge is to turn the global connectedness constraint into a set of local constraints. Here we describe a local-interaction model that achieves this objective. Our model leads to a coarse-grained message-passing algorithm involving only five coarse-grained states for each vertex, after a simplifying process of state aggregation.

\section{The spin glass model}
\label{sec:model}

Following earlier studies on the feedback vertex set problem~\cite{Zhou-2013,Qin-2018}, the minimum $K$-core attack problem~\cite{Zhou-2022} and also the Steiner tree problem~\cite{Bayati-etal-2008}, we assign to each vertex $i$ a discrete state $A_i$, which can take $(d_i+2)$ possible values. Two of these values (denoted as $0$ and $0^*$) correspond to the situation of vertex $i$ not belonging to the backbone $\Gamma$, and the remaining $d_i$ values with $A_i \in \partial i$ correspond to the complementary situation of vertex $i$ belonging to the backbone $\Gamma$. We impose a set of local constraints concerning the state $A_i$ of vertex $i$ and those of its nearest neighbors:
\begin{enumerate}
    \item[1.] If $A_i = 0$ (the normal empty state), then the empty vertex $i$ must have two or more occupied nearest neighbors.
    \item[2.] If $A_i = 0^*$ (the critically empty state), then the empty vertex $i$ must only have one occupied nearest neighbor.
    \item [3.] If $A_i = j$, then $j$ must be the index of a vertex in the neighborhood of the occupied vertex $i$ (namely, $j \in \partial i$), and vertex $j$ must also be occupied (that is, $A_j \neq 0, 0^*$).
\end{enumerate}
We distinguish between the empty states $0$ and $0^*$ purely for the convenience of writing down the cavity equations (see Appendix~~\ref{app:RSeqs}). State $0$ means that vertex $i$ can be accessed through two or more neighboring vertices in the backbone, while state $0^*$ indicates that vertex $i$ can be accessed only through one neighbor in the backbone (so this occupied neighbor can not leave the backbone).  As a consequence of the third local rule, the occupied vertices will organize into some connected subgraphs, each of which contains at least one loop but usually many loops. A visual indication of the state $A_i = j$ is to decorate the edge $(i, j)$ by an arrowhead pointing from vertex $i$ to vertex $j$. In later discussions when we say `$i$ points to $j$' we mean that $A_i = j$.  To encourage the formation of a single connected component of occupied vertices instead of many such components, we introduce an additional rule:
\begin{enumerate}
    \item[4.] If vertex $i$ is occupied ($A_i \neq 0, 0^*$), then vertex $i$ must be pointed to by at least one occupied vertex or it has at least one critically empty neighbor.
\end{enumerate}
\begin{figure}
\centering
\subfigure[]{
\includegraphics[width=0.25\linewidth]{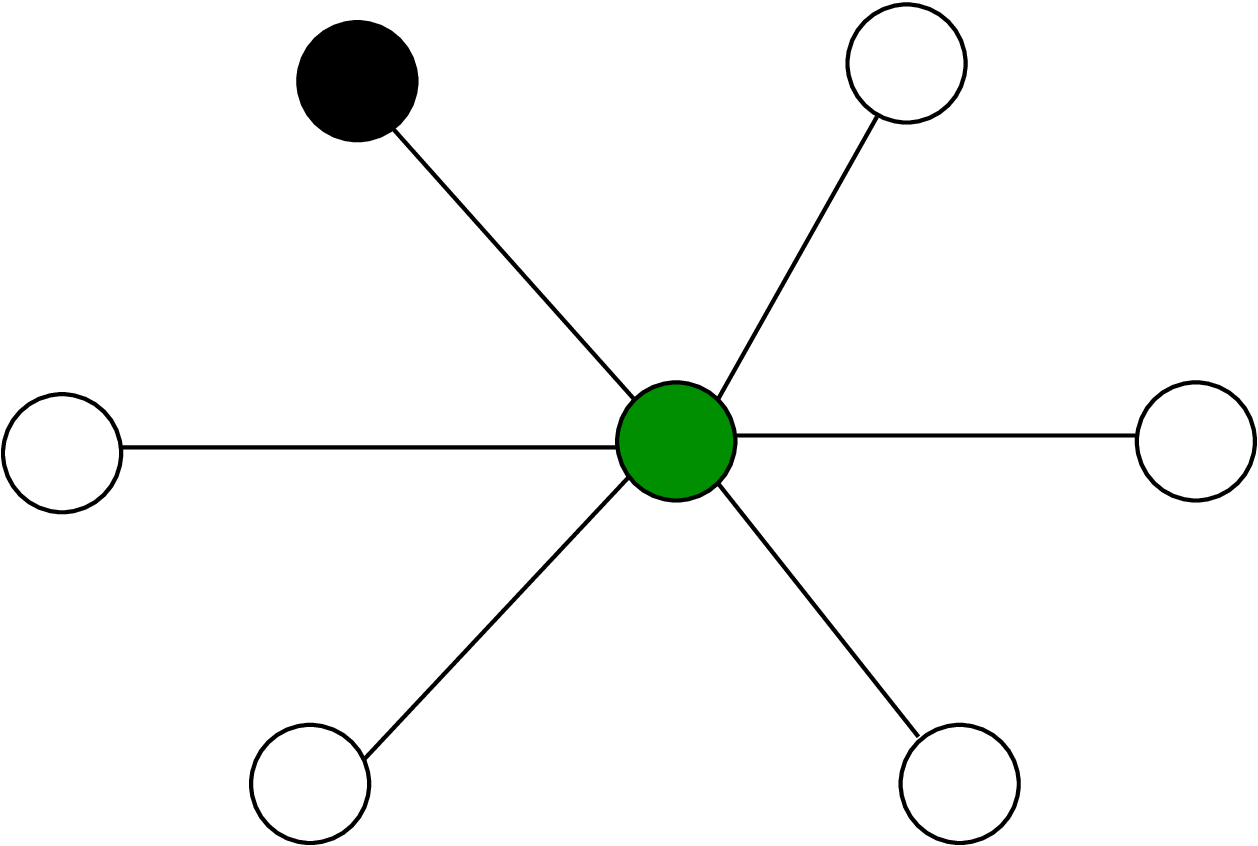}
}
\subfigure[]{
\includegraphics[width=0.25\linewidth]{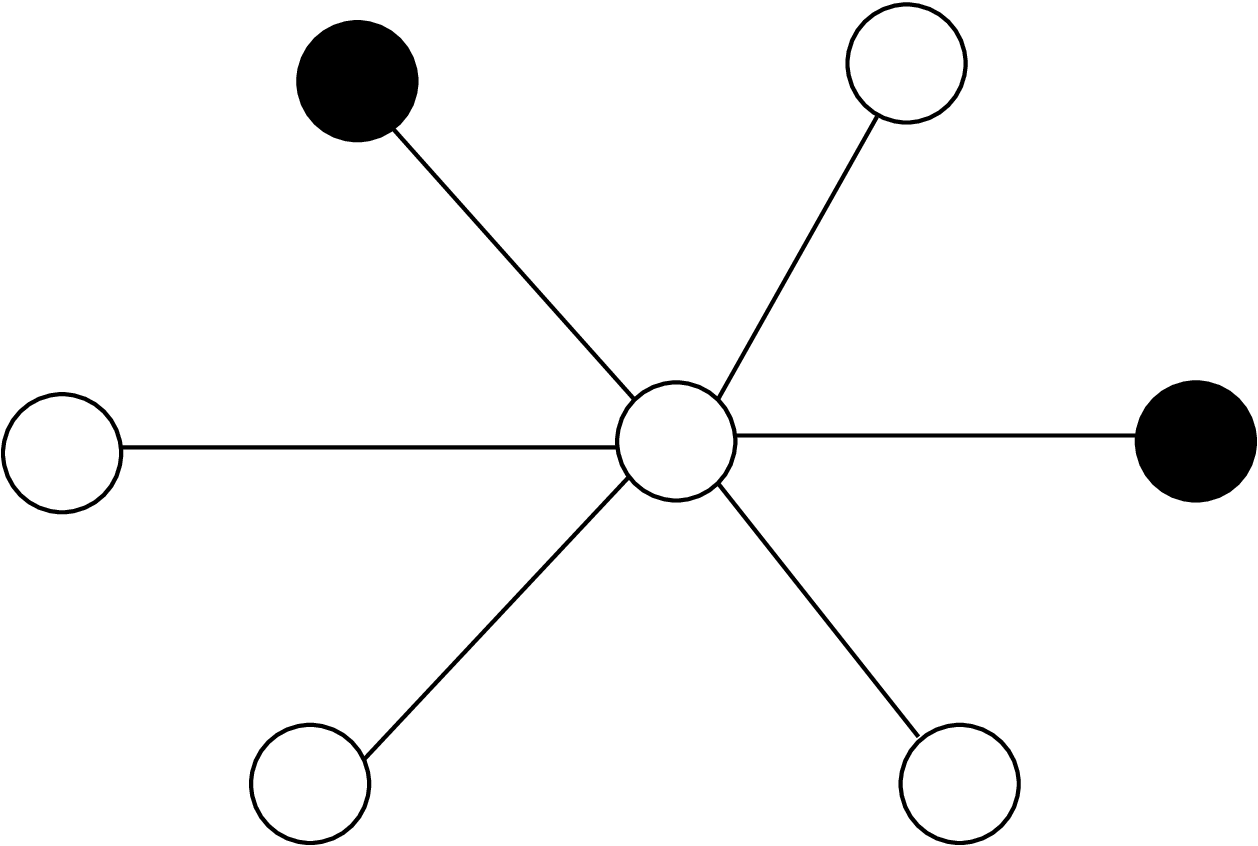}
}
  
\subfigure[]{
\includegraphics[width=0.25\linewidth]{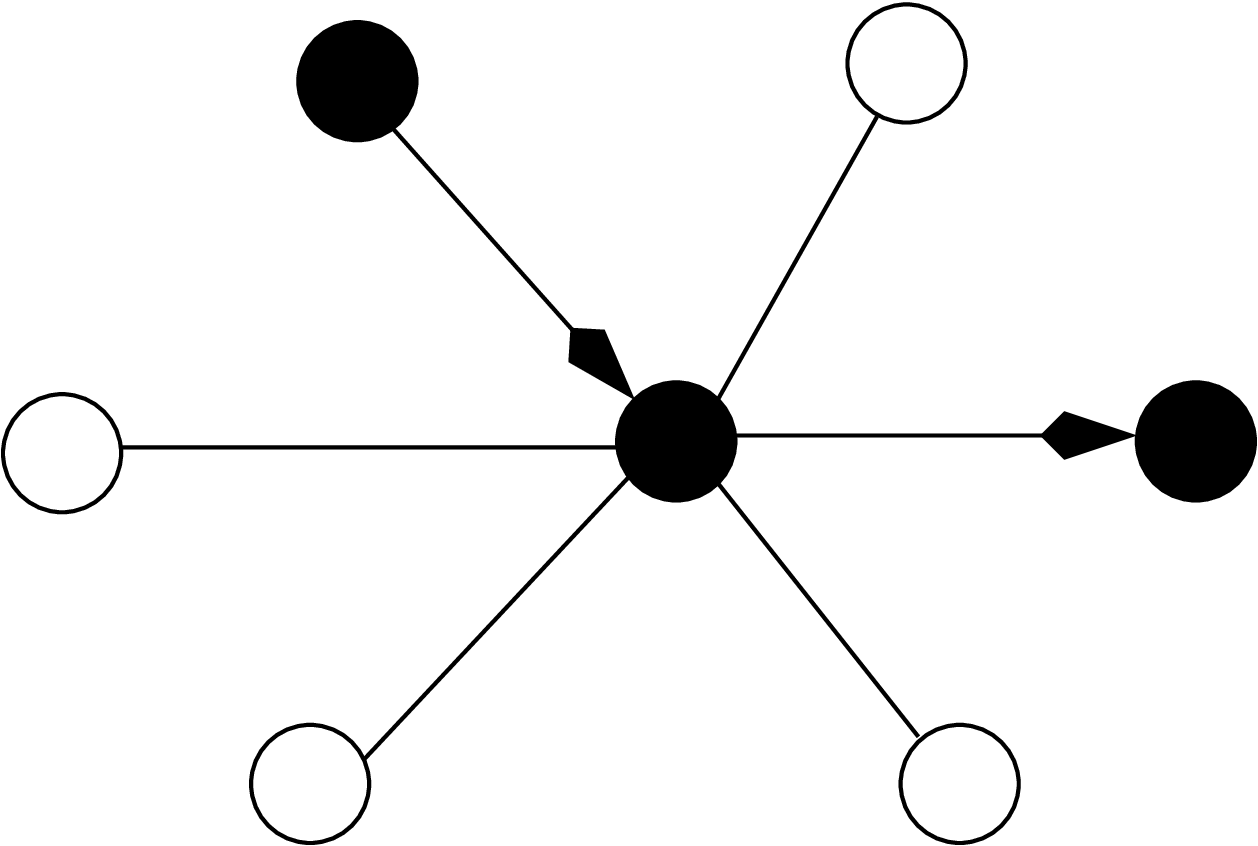}
} 
\subfigure[]{
\includegraphics[width=0.25\linewidth]{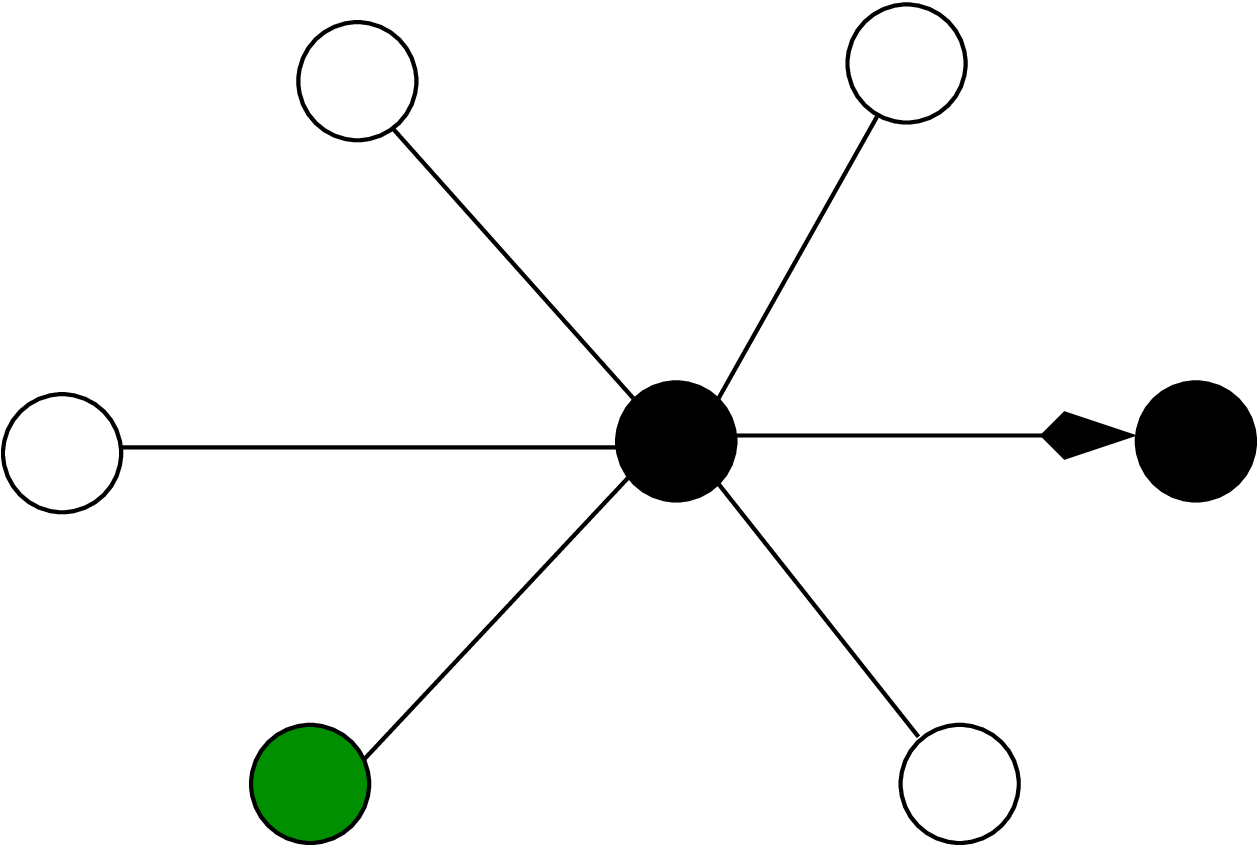}
}
\caption{The local constraints of our model. Black circles indicate occupied vertices, white and green circles indicate empty vertices. (a) Critically empty vertex only has one occupied neighbor. (b) Normal empty vertex has at least two occupied neighbors. (c) If no neighbor of an occupied node is a critical empty node, then at least one occupied neighbor node points to it. (d) If an occupied vertex has a critically empty neighbor, then it does not need to be pointed to by any occupied vertex.
}
\label{fig:localrule}
\end{figure}

The local constraints of our model promote the occupied nodes to form a connected subgraph. First, each occupied vertex $i$ is required to point to another occupied vertex $j$; second, if vertex $i$ points to $j$, then vertex $j$ can not point to $i$ but has to point to another of its occupied neighbors; third, each occupied vertex $i$ must be pointed to by an occupied neighbor (except for the special case of $i$ having a critically empty neighbor). In the rare situation of the occupied vertices form two or more disconnected components, it will be easy to connect these components into a single connected component by adding a few empty vertices to the list of occupied vertices.

From the definition of vertex state it is obvious that each node $i$ brings a constraint to its own state $A_i$ and also to the states of all its nearest neighbors (Fig.~\ref{fig:localrule}). To signify whether or not the local constraints at vertex $i$ are satisfied, we define a constraint factor $C_i \in \{0, 1\}$ as
\begin{equation}
\begin{split}
& C_{i}(A_{i},\underline{A}_{\partial i}) \ = \  \delta_{A_i}^{0^{*}} I\Bigl( \sum_{k \in \partial  i} \sum\limits_{m\in \partial k\backslash i} \delta_{A_{k}}^m -  1 \Bigr) \prod_{j \in \partial  i} (1 - \delta_{A_j}^i ) 
\\
& \quad + \delta_{A_i}^{0} \Theta\Bigl( \sum_{k \in \partial i} \sum\limits_{m\in \partial k \backslash i} \delta_{A_k}^m - 2 \Bigr) \prod_{j \in \partial i} ( 1 - \delta_{A_j}^i ) \\
& \quad + \sum_{j\in\partial i} \delta_{A_i}^j  \sum\limits_{k\in \partial j\backslash i} \delta_{A_j}^k \Theta \Bigl( \sum_{m \in \partial i \backslash j} \bigl( \delta_{A_m}^i + \delta_{A_m}^{0^*} \bigr) - 1 \Bigr) \; .
\end{split}
\label{eq:Ci}
\end{equation}
Here $A_{\partial i} \equiv \{A_j: j\in \partial i\}$ denotes the composite state of the neighboring vertices of vertex $i$;  $\delta_x^y$ is the Kronecker symbol such that $\delta_x^y=1$ if $x$ and $y$ are identical and $\delta_x^y=0$ if $x$ and $y$ are different; $I(x)$ is the indicator function such that $I(x)=1$ if $x=0$ and $I(x)=0$ if $x\neq 0$; and  $\Theta(x)$ is the Heaviside step function whose value is $\Theta(x) = 1$ for $x\geq 0$ and $\Theta(x) = 0$ for $x < 0$.  The last sum means that, if vertex $i$ points to vertex $j$, then vertex $j$ must be pointing to one of its nearest neighbors (say vertex $k$) other than $i$ and that at least one of the nearest neighbors of vertex $i$ (say vertex $m$) must either be pointing to $i$ ($A_m=i$) or be critically empty ($A_m = 0^*$).

Notice that the local constraint associated with vertex $i$ is violated if $C_i=0$ and it is satisfied if $C_i=1$. For a microscopic configuration $(A_1, \ldots, A_N)$ to be a valid one, it must satisfy all the vertex constraints, which means $C_i = 1$ for all the vertices. The partition function of the model is a weighted sum over all the possible microscopic configurations:
\begin{equation}
\label{eq:Zbeta}
Z(\beta) = \sum_{A_{1}, \ldots , A_{N}}\; \prod_{i=1}^{N} \biggl[ e^{ -\beta (1- \delta_{A_{i}}^{0}-\delta_{A_{i}}^{0^*})} C_{i}(A_{i},\underline{A}_{\partial i}) \biggr] \; ,
\end{equation}
where $\beta > 0$ is the inverse temperature parameter. If a vertex $i$ is occupied ($A_i \neq 0, 0^*$) then there is a penalty factor $e^{-\beta}$. Because of the vertex factors $C_i$ of Eq.~(\ref{eq:Zbeta}), it is clear that only those microscopic configurations which satisfy all the local constraints have non-vanishing contributions to $Z(\beta)$. When $\beta$ becomes sufficiently large, the partition function will be predominantly contributed by those valid configurations which contain the minimum number of occupied vertices.

The spin glass model contains only local interactions and it can be studied by many different methods, such as simulated annealing or other heuristic optimization algorithms. Here we solve the spin glass model Eq.~(\ref{eq:Zbeta}) by the standard cavity method of statistical mechanics. The belief propagation (BP) equations are derived by similar method as that of Ref.~\cite{Zhou-2022}. We present the essential technical details in Appendix~\ref{app:RSeqs}. The BP equations at the coarse-grained level are Eqs.~(\ref{eq:Qji0})--(\ref{eq:Qji4}), which only involve five coarse-grained states. Therefore our model is essentially a five-state Potts model with local interactions for an optimization problem with a global connectivity constraint. We can iterate these coarse-grained BP equations on single graph instances or through population dynamics on a given ensemble of random graphs~\cite{Zhao-Habibulla-Zhou-2015}. 

The next two sections will discuss the results obtained by this mean field theory.

\section{Regular random ensembles}

\begin{figure}[t]
  \centering
  \subfigure[]{
  \includegraphics[width=0.35\linewidth]{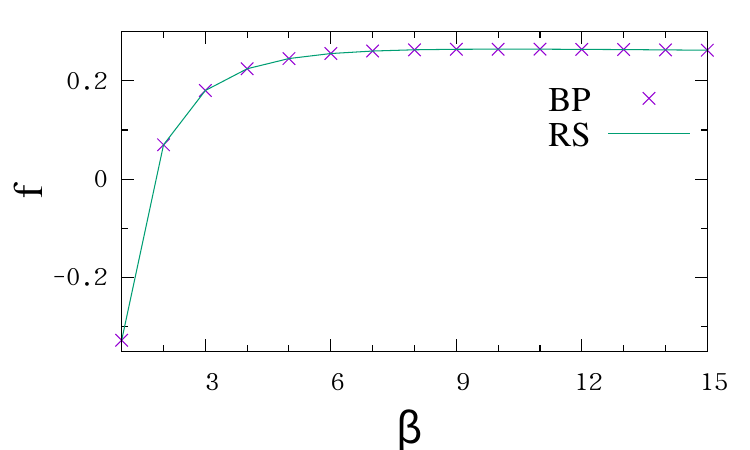}
  }
   \subfigure[]{
  \includegraphics[width=0.35\linewidth]{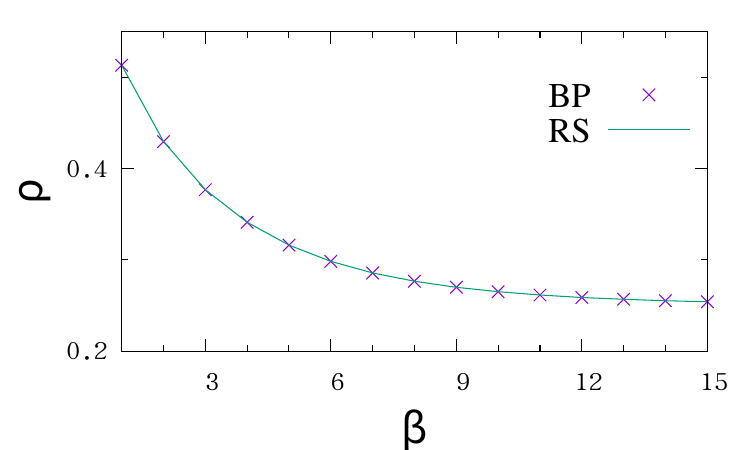}
  }
   \subfigure[]{
  \includegraphics[width=0.35\linewidth]{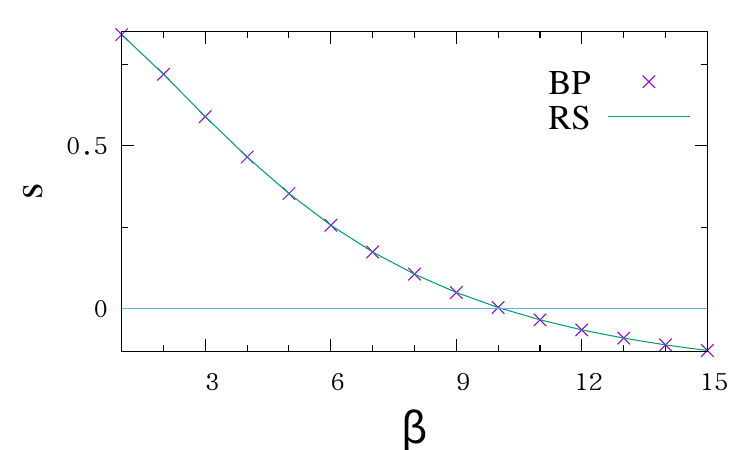}
  } 
  \subfigure[]{
  \label{fig:RRk5:svsr}
  \includegraphics[width=0.35\linewidth]{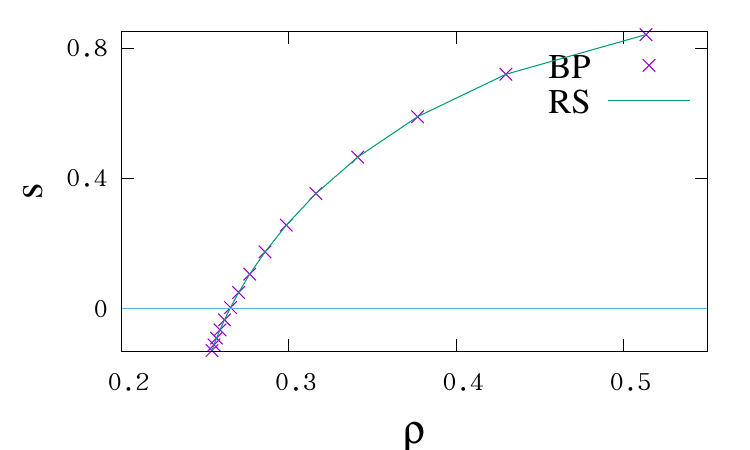}
  }
  \caption{
  Replica-symmetric population dynamics simulation results on the enesmble of random regular networks of integer degree $K = 5$ (RS, symbols), and results obtained by belief-propagation iterations on a single RR network instance of size $N=10^4$ and the same degree $K=5$ (BP, lines). (a) Free energy density $f$ versus inverse temperature $\beta$; (b) energy density $\rho$ versus $\beta$; (c) entropy density $s$ versus $\beta$; and (d) entropy density $s$ versus energy density $\rho$. 
}
  \label{fig:RRk5}
\end{figure}

Each vertex has $K$ nearest neighbors in a regular random (RR) graph, which are chosen completely at random among all the vertices of the graph. We apply the replica-symmetric (RS) mean field theory both to the whole ensemble of RR graphs with degree $K$ and to single RR graph instances. As a concrete example, we show in Fig.~\ref{fig:RRk5} the theoretical results obtained at $K=5$ for the graph ensemble (cross symbols) and for a single graph instance (solid line). For this degree $K=5$, the ensemble-averaged RS results are indistinguishable from the single-instance BP results. The energy density $\rho$ and the entropy density $s$ both decrease with inverse temperature $\beta$, as expected. Negative values of $s$ are predicted for $\beta > 10.1$, indicating failure of the RS mean field theory at high $\beta$ values. Within the framework of the RS mean field theory we take the energy density corresponding to zero entropy density [Fig.~\ref{fig:RRk5:svsr}] as the minimum energy density $\rho_0$ of the network ensemble. For $K=5$ the value is $\rho_0 = 0.265$, which means that the minimum backbone contains at least $0.265 N$ vertices of the random graph.

\begin{table}
    \caption{
    Relative size $\rho_0$ of minimum backbone (or connected dominating set) of the regular random network ensemble with degree $K$. Results are obtained by the replica-symmetric mean field theory. As comparison we also list the relative size $\rho_0^{DS}$ of minimum dominating set as obtained in Ref.~\cite{Zhao-Habibulla-Zhou-2015}.
    }
    \centering 
    \begin{tabular}{lll|lll}
    $K$\quad\quad & $\rho_0^{DS}$\quad \quad \quad & $\rho_0$\quad \quad \quad & \;  $K$ \quad \quad & $\rho_0^{DS}$ \quad \quad \quad & $\rho_0$\quad \quad  \\
    \hline 
      $3$  & $0.2637$ & $0.4922$ &\; $12$ & $0.1097$ & $0.1206$\\
      $4$  & $0.2225$ & $0.3370$ &\; $13$ & $0.1039$ & $0.1130$ \\
      $5$  & $0.1941$ & $0.2646$ &\; $14$ & $0.0987$ & $0.1062$ \\
      $6$  & $0.1731$ & $0.2213$ &\; $15$ & $0.0941$ & $0.1006$ \\
      $7$  & $0.1569$ & $0.1919$ &\; $16$ & $0.0900$ & $0.0955$ \\
      $8$  & $0.1438$ & $0.1703$ &\; $17$ & $0.0863$ & $0.0909$ \\
      $9$  & $0.1331$ & $0.1537$ &\; $18$ & $0.0829$ & $0.0869$ \\
      $10$ & $0.1241$ & $0.1406$ &\; $19$ & $0.0798$ & $0.0832$\\
      $11$ & $0.1164$ & $0.1297$ &\; $20$ & $0.0770$ & $0.0799$\\
      \hline
        \end{tabular}
    \label{tab:RRmin}
\end{table}

The predicted minimum relative sizes $\rho_0$ of backbones for RR network ensembles of degree $K$ up to $K=20$ are listed in Tab.~\ref{tab:RRmin} (the column for backbone, $\rho_{0}$). As expected, the value of $\rho_0$ decreases with vertex degree $K$. We also copy the predicted minimum relative sizes of dominating set as obtained in Ref.~\cite{Zhao-Habibulla-Zhou-2015} to this table (the column for dominating set, $\rho_0^{DS}$). When $K$ increases the gap between these two minimum relative sizes shrinks, indicating that the connectivity constraint is less significant for denser network ensembles. This may have algorithmic consequences. Tab.~\ref{tab:RRmin} also indicates that, when the degree $K$ becomes large, the relative size of minimum dominating sets could serve as a good lower-bound for the relative size of minimum backbones. 

For single random graph instances with degree $K\leq 8$, we find the coarse-grained BP iterations are able to converge even at high values of $\beta$ at which the entropy density values become slightly negative. When $ K \geq 9$, the coarse-grained BP iterations no longer converge when $\beta$ becomes high and the entropy density $s$ becomes close to zero (for example, at $K = 10$ the BP iteration is non-convergent for $\beta > 8.1$). The non-convergence of BP is another strong indication of the inadequacy of the RS mean field theory at high $\beta$ values. We expect that the system will enter the spin glass phase at certain critical value of $\beta$, whose value may be precisely predictable following the same theoretical framework adopted in Ref.~\cite{Qin-etal-2016}.  Such an analysis will be reported in a separate paper. 

\section{Erd\"os-R\'enyi ensembles}

An Erd\"os-R\'enyi (ER) random graph is constructed by choosing $M$ edges out of the $N (N-1)/2$ possible edges uniformly at random and adding them to the graph. The mean vertex degree of the graph is then simply $2 M / N$. There may be isolated vertices in such a random graph. To guarantee that the whole random graph is connected, in the present work we first assign to each vertex $i$ a degree $d_i \geq k_0$, following the following modified Poisson distribution,
\begin{equation}
P(d_i) = \frac{1}{ \sum_{d \geq k_0} e^{-c} c^d / d!} \frac{ e^{-c} c^{d_i}}{d_i !} \; ,
\label{eq:ERpd}
\end{equation}
where $c$ is the mean degree parameter and $k_0$ is the minimum vertex degree (we set $k_0=4$). Then we add edges uniformly at random between any two vertices $i$ and $j$ until all the vertices $i$ are connected to $d_i$ other vertices. After this initial graph is constructed, we then further randomize the graph to remove structural correlations. The connectivity of the final random graph is checked. The mean vertex degree of such a random ER graph   is
\begin{equation}
\label{eq:ermd}
\langle d \rangle = \frac{ \sum_{d_i \geq k_0} d_i\; e^{-c} c^{d_i} / d_i !}{ \sum_{d \geq k_0} e^{-c} c^d / d!} \; .
\end{equation}

\begin{table}
    \caption{
    Relative size $\rho_0$ of minimum backbone (or connected dominating set) of the Erd\"os-R\'enyi random network ensemble with minimum degree $k_0 = 4$ and parameter $c$ (the mean degree $\langle d \rangle$ is determined by Eq.~(\ref{eq:ermd})). Results are obtained by the replica-symmetric mean field theory.
    }
    \centering 
    \begin{tabular}{lll|lll|lll}
    $c$ \quad\quad & $\langle d \rangle$ \quad \quad & $\rho_0$ \quad \quad & \; $c$ \quad \quad & $\langle d \rangle$ \quad \quad & $\rho_0$ 
    & \;  $c$ \quad \quad & $\langle d \rangle$ \quad \quad & $\rho_0$ \quad \\
    \hline 
      $1.0$ \quad \quad & $4.2288$ \quad \quad & $0.3018$ \quad   & \;
      $7.5$ \quad \quad & $7.7558$ \quad \quad & $0.1548$ \quad & \;  
      $14.0$ \quad \quad & $14.009$ \quad \quad & $0.0953$ 
      \\
      \hline
      $1.5$ \quad \quad & $4.3788$ \quad \quad & $0.2852$ \quad & \; 
      $8.0$ \quad \quad & $8.2444$ \quad \quad & $0.1470$ \quad & \; 
       $14.5$ \quad \quad & $14.521$ \quad \quad & $0.0926$ 
       \\
      \hline
      $2.0$ \quad \quad & $4.5322$ \quad \quad & $0.2711$ \quad & \;
       $8.5$ \quad \quad & $8.691$ \quad \quad & $0.1409$ \quad & \; 
       $15.0$ \quad \quad & $15.0358$ \quad \quad & $0.0903$ 
      \\
      \hline
      $2.5$ \quad \quad & $4.7166$ \quad \quad & $0.2564$ \quad & \;
      $9.0$ \quad \quad & $9.1392$ \quad \quad & $0.1347$ \quad &  \;
       $15.5$ \quad \quad & $15.466$ \quad \quad & $0.0879$ 
     \\
      \hline
      $3.0$ \quad \quad & $4.9174$ \quad \quad & $0.2432$ \quad & \;
      $9.5$ \quad \quad & $9.6232$ \quad \quad & $0.1294$ \quad & \;
       $16.0$ \quad \quad & $15.932$ \quad \quad & $0.0861$
     \\
      \hline
      $3.5$ \quad \quad & $5.1172$ \quad \quad & $0.2313$ \quad & \;
      $10.0$ \quad \quad & $10.0994$ \quad \quad & $0.1240$ \quad & \; 
       $16.5$ \quad \quad & $16.512$ \quad \quad & $0.0834$ 
      \\
      \hline
      $4.0$ \quad \quad & $5.3544$ \quad \quad & $0.2199$ \quad & \;
      $10.5$ \quad\quad & $10.5518$ \quad\quad & 0.1196 & \; 
       $17.0$ \quad \quad & $16.9788$ \quad \quad & $0.0819$
      \\
      \hline
      $4.5$ \quad \quad & $5.6526$ \quad \quad & $0.2072$ \quad & \;
       $11.0$ \quad \quad & $11.0276$ \quad \quad & $0.1158$ & \;
     $17.5$ \quad \quad & $17.5514$ \quad \quad & $0.0799$
     \\
      \hline
      $5.0$ \quad \quad & $5.9554$ \quad \quad & $0.1965$ \quad & \;
      $11.5$ \quad \quad & $11.5496$ \quad \quad & $0.1115$ & \; 
      $18.0$ \quad \quad & $18.0822$ \quad \quad & $0.0782$
      \\
      \hline
      $5.5$ \quad \quad & $6.282$ \quad \quad & $0.1864$ \quad & \;
      $12.0$ \quad \quad & $11.9724$ \quad \quad & $0.1081$ & \; 
      $18.5$ \quad \quad & $18.4754$ \quad \quad & $0.0767$ 
     \\
      \hline
      $6.0$ \quad \quad & $6.6304$ \quad \quad & $0.1778$ \quad & \;
       $12.5$ \quad \quad & $12.459$ \quad \quad & $0.1048$ & \; 
       $19.0$ \quad \quad & $19.0438$ \quad \quad & $0.0749$
      \\
      \hline
      $6.5$ \quad \quad & $6.9926$ \quad \quad & $0.1694$ \quad & \;
       $13.0$ \quad \quad & $12.99$ \quad \quad & $0.1012$ & \; 
       $19.5$ \quad \quad & $19.5372$ \quad \quad & $0.0734$
     \\
      \hline
      $7.0$ \quad \quad & $7.389$ \quad \quad & $0.1614$ \quad & \;
      $13.5$ \quad \quad & $13.487$ \quad \quad & $0.0982$ & \; 
      $20.0$ \quad \quad & $19.9484$ \quad \quad & $0.0723$
      \\
\hline
        \end{tabular}
    \label{tab:ERmin}
\end{table}

We carry out RS population dynamics simulations on the ER ensembles with given degree parameter $c$ (and $k_0=4$) and BP iterations on single ER graph instances. In the RS population dynamics, the input is a single vertex degree array containing a large number ($=10^4$) vertex degrees $d_i$ sampled according to Eq.~(\ref{eq:ERpd}), see Table~\ref{tab:ERmin} for the list of $c$ and mean degree $\langle d \rangle$ used in the RS population dynamics simulations.

\begin{figure}[]
  \centering
  \subfigure[]{
  \includegraphics[width=0.35\linewidth]{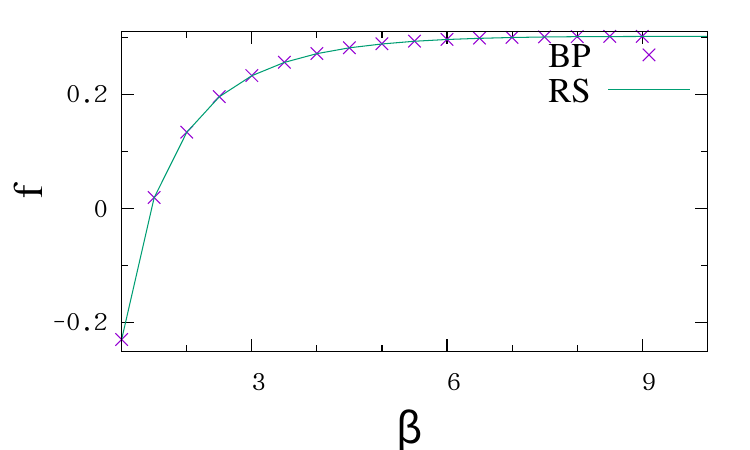}
  }
   \subfigure[]{
  \includegraphics[width=0.35\linewidth]{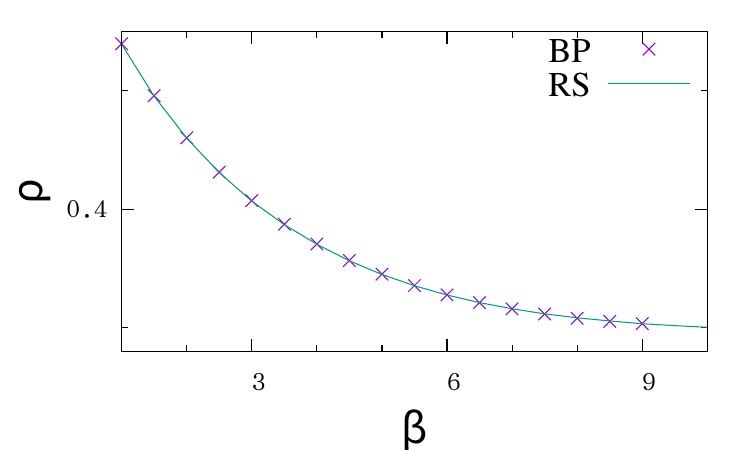}
  }
\subfigure[]{
 \includegraphics[width=0.35\linewidth]{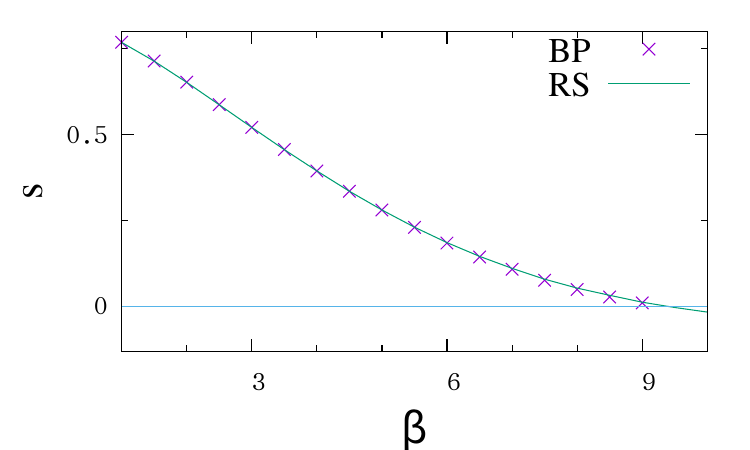}
 } 
 \subfigure[]{
 \includegraphics[width=0.35\linewidth]{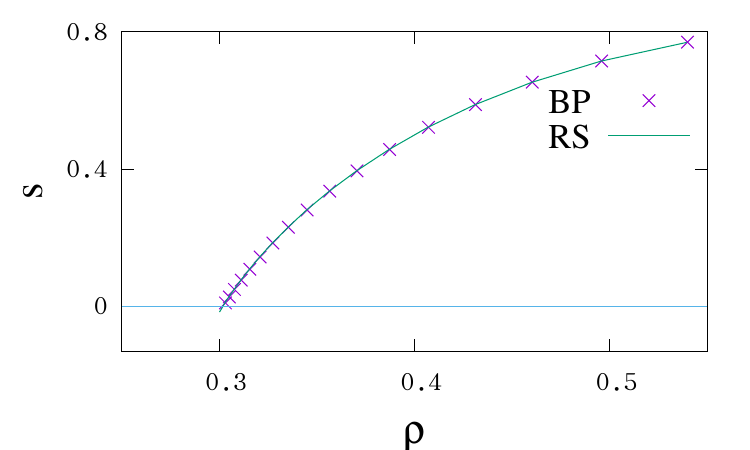}
 }
 \caption{Same as Fig.~\ref{fig:RRk5}, but for Erd\"os-R\'enyi random graphs  with structural parameters $c=1$ and $k_0=4$, with mean vertex degree $\langle d \rangle = 4.2288$.
 }
 \label{fig:ERc1k4}
\end{figure}

Figure~\ref{fig:ERc1k4} compares the ensemble-averaged and single instance results for the case of $c = 1$ and $k_0 = 4$, with mean vertex degree is $\langle d \rangle \approx 4.229$. When the BP iteration is convergent (this occurs for $\beta < 8.7$) we find that the single-graph results are the same with the ensemble-averaged results. We again take the minimum energy density $\rho_0$ by looking for the point of zero entropy density of the RS results (which is $\rho_0 = 0.3018$). We summarize in Table~\ref{tab:ERmin} our theoretical results on the ER graph ensembles. These theoretical results will be compared with algorithmic results in the following Fig.~\ref{fig:bbqER}.

Similar to the situation encountered for the RR graph cases, the coarse-grained BP iterations are not convergent at high values of $\beta$. This indicates the existence of low-energy spin glass states. For a given ER or RR random graph ensemble, the predicted minimum energy density $\rho_0$ by the RS mean field theory is very likely only a lower bound of the true minimum energy density. It is necessary to study the model in the future at the level of first-order replica-symmetry breaking to improve the theoretical results on the minimum energy density.

\section{Message-passing algorithm for single graph instances}

Some of the widely known heuristic algorithms for the minimum CDS problem were based on simple greedy strategies, such as choosing vertices according to their degrees~\cite{GuhaKhuller1998}. Inspired by the backbone model and the coarse-grained BP equations, we implement a message-passing heuristic algorithm, {\tt BackBoneQueue} ({\tt BBQ}), to build an approximately minimum-sized backbone $\mathcal{B}$ and the associated connected dominating set $\Gamma$ for an input random graph $G$. Initially the CDS $\Gamma$ is empty and all the $N$ vertices of graph $G$ are queued as candidate members of $\Gamma$. At each step of the {\tt BBQ} process we add a tiny fraction of the vertices in this queue to the CDS until there is no need to expand this set $\Gamma$ further. The final step of {\tt BBQ} is to check the global connectivity of the backbone $\mathcal{B}$. If it is not yet connected, we add a few extra vertices to make the backbone to be connected.

At the start of the {\tt BBQ} process all the vertices $j$ of the input graph $G$ are assigned the empty state $A_j=0$. Then at each elementary step of the {\tt BBQ} process, we update the coarse-grained cavity probabilities $(Q_{j\rightarrow i}^0, Q_{j\rightarrow i}^1, Q_{j\rightarrow i}^2, Q_{j\rightarrow i}^3, Q_{j\rightarrow i}^4)$ between pairs of empty vertices $j$ and $i$ in a random sequential order. If all the adjacent vertices of the focal empty vertex $j$ are all empty, then the cavity probability values $Q_{j\rightarrow i}^{s}$ (with $s=0, \ldots, 4$) from $j$ to an adjacent empty vertex $i$ are updated according to Eqs.~(\ref{eq:Qji0})--(\ref{eq:Qji4}). On the other hand if the focal empty vertex $j$ has one or more occupied nearest neighbors, then the cavity probability values $Q_{j\rightarrow i}^s$ from $j$ to an adjacent empty vertex $i$ are updated according to Eqs.~(\ref{eq:Qji0mod1})--(\ref{eq:Qji4mod1}). We emphasize that the set $\partial j$ in these message passing equations (\ref{eq:Qji0mod1})--(\ref{eq:Qji4mod1}) only contains the empty nearest neighbors of vertex $j$, namely, $\partial j = \{i: (i, j) \in G, A_i=0\}$.

We run the coarse-grained BP equations (\ref{eq:Qji0})--(\ref{eq:Qji4}) and (\ref{eq:Qji0mod1})--(\ref{eq:Qji4mod1}) for the subgraph of $G$ containing all the empty vertices for a small number of times (say $10$ repeats). Then we update the marginal probability $Q_j^b$ of an empty vertex $j$ being moved to the backbone as follows: If vertex $j$ has no occupied nearest neighbors, this probability is updated according to Eq.~(\ref{eq:Qjb}); if vertex $j$ has at least one occupied nearest neighbors, then this probability is updated according to Eq.~(\ref{eq:Qjbmod1}). We then queue all the empty vertices $j$ in descending order of their $Q_j^b$ values, and then move a small fraction (say $0.01$) of the empty vertices at the top of this queue to the nascent dominating set $\Gamma$ and update the corresponding backbone $\mathcal{B}$.

If all the vertices of the input graph $G$ are either labeled as being occupied or are connected to at least one occupied nearest neighbors, then the coarse-grained BP iteration and decimation process will not be continued. We then make a final check on the connection property of the backbone $\mathcal{B}$. If this backbone is formed by two or more disconnected components, then we search for a shortest path in the original graph $G$ which is connecting two connected components of the backbone and all the vertices of this shortest path to the backbone. This merging process is continued after the whole backbone is a single connected subgraph. The CDS associated with this backbone $\mathcal{B}$ is then considered an output CDS of this whole run of the {\tt BBQ} dynamical process.

\begin{figure}
  \centering
  \subfigure[]{
  \includegraphics[width=0.45\linewidth]{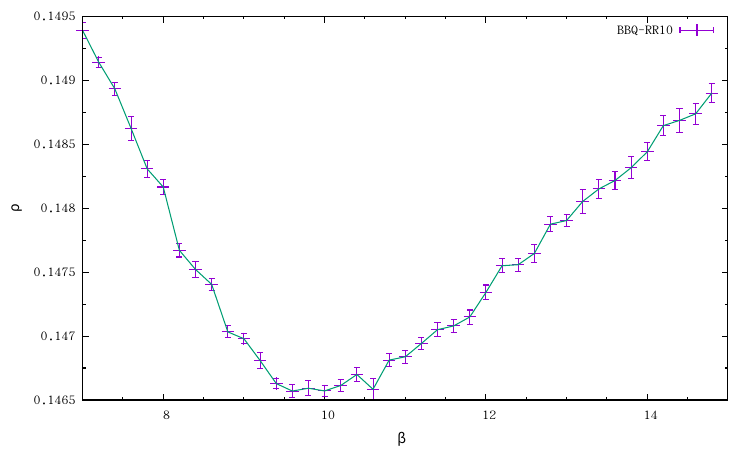}
  }
\subfigure[]{
    \includegraphics[width=0.45\linewidth]{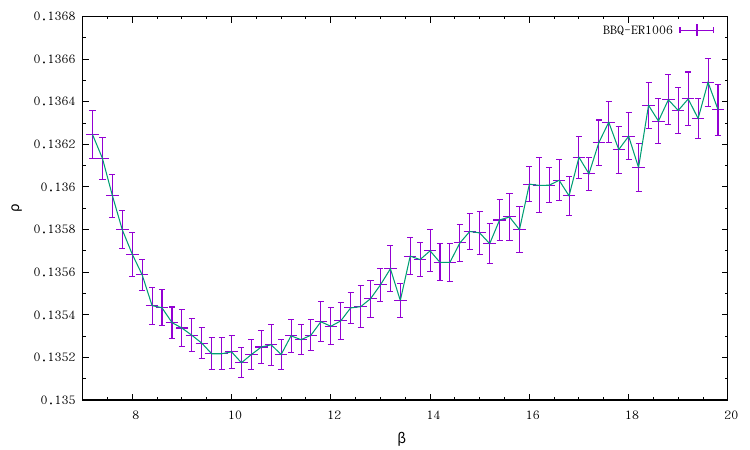}
    }
  \caption{
  The effect of the inverse temperature $\beta$ to the performance of the {\tt BBQ} algorithm. (a) Average results obtained on $24$ regular random graph instances containing $N=10000$ vertices and vertex degree $K=10$. (b) Average results obtained on $24$ Erd\"os-R\'enyi random graph instances of size $N=10000$ and mean degree $\langle d \rangle = 10.06$.
  }
\label{fig:bbqBeta}
\end{figure}

An important parameter of the ${\tt BBQ}$ algorithm is the inverse temperature $\beta$. We need to choose a large value of $\beta$ to encourage more vertices to be empty, but on the other hand, if $\beta$ is too large the system will be deep in the spin glass phase and the replica-symmetric mean field theory will no longer be valid. The optimal value of $\beta$ may be different for different random graph ensembles. For the RR and ER random graphs with mean degree $\langle d \rangle \approx 10$, our empirical results suggest that $\beta = 10$ is a good choice (Fig.~\ref{fig:bbqBeta}). In the present work we then simply fix $\beta=10$ for the ${\tt BBQ}$ algorithm for all the random graph instances whose mean degrees ranging from $4$ to $20$.

\begin{figure}
  \centering
  \subfigure[]{
  \label{fig:bbqRR}
  \includegraphics[width=0.45\linewidth]{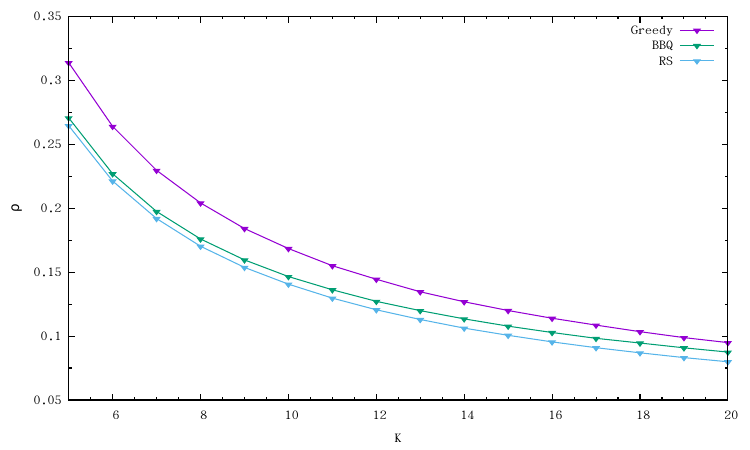}
  }
\subfigure[]{
\label{fig:bbqER}
    \includegraphics[width=0.45\linewidth]{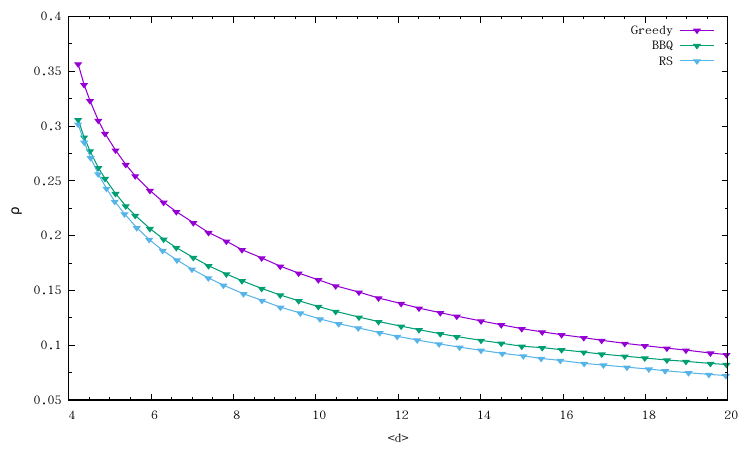}
    }
  \caption{
  Performance of the {\tt BBQ} message-passing algorithm on regular random graphs (a) and Erd\"os-R\'enyi random graphs (b), and the comparison with the theoretical-prediction of the replica-symmetric (RS) mean-field theory and with the modified greedy algorithm of Ref.~\cite{GuhaKhuller1998}. The energy density $\rho$ at each value of degree $K$ (for RR graphs) or mean degree $\langle d\rangle$ (for ER graphs) are the average value over $24$ graph instances of size $N=10000$. The inverse temperature parameter of the {\tt BBQ} algorithm is fixed at $\beta=10$. The {\tt BBQ} and greedy algorithms run on each graph instance only once. The theoretical RS energy densities $\rho$ are obtained by population dynamics with the vertex degree distributions as inputs (see Table~\ref{tab:ERmin}).}
  \label{fig:bbqcomp}
\end{figure}
  
The numerical results obtained by the {\tt BBQ} algorithm for RR and ER graph instances are summarized in Fig.~\ref{fig:bbqRR} and Fig.~\ref{fig:bbqER}, respectively. We see that the {\tt BBQ} algorithm greatly outperforms the modified greedy algorithm of Ref.~\cite{GuhaKhuller1998} considerably. On the other hand, the {\tt BBQ} algorithmic results are still higher than the predicted minimum energy density values by the RS mean field theory, indicating there is sill much room for improving this algorithm and that the RS prediction also needs to be improved. At the present stage, the comparative results of Fig.~\ref{fig:bbqcomp} suggest that the simple local-interaction model is very useful both for estimating the minimum CDS relative size and for constructing approximately minimum CDS solutions. As we also mentioned earlier, our model is also convenient for running simulated annealing dynamics or other local search algorithms on single graph instances.

\section{Conclusion}

We presented a spin glass model for the minimum backbone problem and the closely related minimum dominating set problem and derived a set of coarse-grained belief propagation equations. The main innovation of this work is that we implemented the global constraint of a connected dominating set through a set of local constraints to avoid the high computational demand of checking for the satisfaction of the global constraint. Based on this essentially $5$-states local-interaction Potts model, we designed a message-passing {\tt BBQ} algorithm to construct close-to-minimum connected dominating sets (and the associated backbone subgraphs) for single random graph instances.

The theoretical and algorithmic results reported in this paper suggested that the local-constraint model is very helpful for tackling the NP-hard minimum backbone problem and the minimum dominating set problem. The same idea probably will also be helpful for some other global constraint problems, such as the travelling salesperson problem and the Hamiltonian cycle problem.

The present work only solved the spin glass model at the replica-symmetric level, completely ignoring the possibility of ergodicity breaking and the emergence of multiple spin glass macroscopic states. Yet the non-convergence of the coarse-grained BP equations at high inverse temperature values strongly indicates that it is necessary to extend the mean field theory at least to the level of first-step replica-symmetry breaking~\cite{Qin-etal-2016,Habibulla-2017}. Much work still needs to be done along this direction, including the determination of the spin glass phase transition points. Besides message-passing algorithms, it is also interesting to solve our local-constraint optimization model on single graph instances by other heuristic algorithms such as simulated annealing.

\begin{acknowledgments}
One of the authors (Y.H.) acknowledges the helpful comments of Haiping Huang and Biaobing Cao. The following funding supports are acknowledged: National Natural Science Foundation of China Grants No. 11421063, No. 11747601 and No.~12247104; National Innovation Institute of Defense Technology Grant No.~22TQ0904ZT01025. Some of the numerical simulations were carried out at the HPC cluster of ITP-CAS and also at the BSCC-A3 platform of the National Supercomputer Center in Beijing.
\end{acknowledgments}

\appendix


\section{The mean field message-passing equations}
\label{app:RSeqs}

We solve the graphical model (\ref{eq:Zbeta}) by the standard replica-symmetric cavity method of statistical mechanics closely following our earlier work~\cite{Zhao-Habibulla-Zhou-2015,Zhou-2022}. Here we write down a set of belief propagation equations for the model (\ref{eq:Zbeta}) and then derive a coarse-grained version of these BP equations. Thermodynamic quantities such as the free energy density are computed using only the coarse-grained BP cavity messages.

\subsection{The belief propagation equations and marginal probabilities}

We define two cavity probabilities $q_{i\rightarrow j}^{A_i, A_j}$ and $q_{j\rightarrow i}^{A_j, A_i}$ on each undirected edge $(i, j)$ of the graph. The quantity $q_{i\rightarrow j}^{A_i, A_j}$ is the joint probability of vertex $i$ being in state $A_i$ and vertex $j$ being in state $A_j$ if the constraint associated with vertex $j$ is lifted but that of vertex $i$ is present. Concerning the cavity probability $q_{j\rightarrow i}^{A_j, A_i}$, if the state of vertex $j$ is taking the empty states $A_j = 0$ or $A_j=0^*$, we have
\begin{eqnarray}
q_{j\rightarrow i}^{0, 0} & \propto & \prod\limits_{k\in \partial j\backslash i} \sum\limits_{A_k} \Bigl[ \delta_{A_k}^0 q_{k\rightarrow j}^{0, 0} + \delta_{A_k}^{0^*} q_{k\rightarrow j}^{0^*, 0} + \sum\limits_{m\in \partial k\backslash  j} \delta_{A_k}^m q_{k\rightarrow j}^{m, 0} \Bigr] \Theta\Bigl( \sum\limits_{l\in \partial j\backslash i} \sum\limits_{n\in \partial l\backslash j} \delta_{A_l}^n - 2 \Bigr) \; , 
\\
q_{j\rightarrow i}^{0, 0^*} & \propto & \prod\limits_{k\in \partial j\backslash i} \sum\limits_{A_k} \Bigl[ \delta_{A_k}^0 q_{k\rightarrow j}^{0, 0} + \delta_{A_k}^{0^*} q_{k\rightarrow j}^{0^*, 0} + \sum\limits_{m\in \partial k\backslash  j} \delta_{A_k}^m q_{k\rightarrow j}^{m, 0} \Bigr] \Theta\Bigl( \sum\limits_{l\in \partial j\backslash i} \sum\limits_{n\in \partial l\backslash j} \delta_{A_l}^n - 2 \Bigr) \; ,
\\
q_{j\rightarrow i}^{0, j} & = & 0 \; , 
\\
q_{j\rightarrow i}^{0, A_i} & \propto & \prod\limits_{k\in \partial j\backslash i} \sum\limits_{A_k} \Bigl[ \delta_{A_k}^0 q_{k\rightarrow j}^{0, 0} + \delta_{A_k}^{0^*} q_{k\rightarrow j}^{0^*, 0} + \sum\limits_{m\in \partial k\backslash  j} \delta_{A_k}^m q_{k\rightarrow j}^{m, 0} \Bigr] \Theta\Bigl( \sum\limits_{l\in \partial j\backslash i} \sum\limits_{n\in \partial l\backslash j} \delta_{A_l}^n - 1 \Bigr) \; , 
\label{eq:qji0Ai}
\\
q_{j\rightarrow i}^{0^*, 0} & \propto & 
\prod\limits_{k\in \partial j\backslash i} \sum\limits_{A_k} \Bigl[ \delta_{A_k}^0 q_{k\rightarrow j}^{0, 0^*} + \delta_{A_k}^{0^*} q_{k\rightarrow j}^{0^*, 0^*} + \sum\limits_{m\in \partial k\backslash  j} \delta_{A_k}^m q_{k\rightarrow j}^{m, 0^*} \Bigr] I\Bigl( \sum\limits_{l\in \partial j\backslash i} \sum\limits_{n\in \partial l\backslash j} \delta_{A_l}^n - 1 \Bigr) \; , 
\\
q_{j\rightarrow i}^{0^*, 0^*} & \propto & \prod\limits_{k\in \partial j\backslash i} \sum\limits_{A_k} \Bigl[ \delta_{A_k}^0 q_{k\rightarrow j}^{0, 0^*} + \delta_{A_k}^{0^*} q_{k\rightarrow j}^{0^*, 0^*} + \sum\limits_{m\in \partial k\backslash  j} \delta_{A_k}^m q_{k\rightarrow j}^{m, 0^*} \Bigr] I\Bigl( \sum\limits_{l\in \partial j\backslash i} \sum\limits_{n\in \partial l\backslash j} \delta_{A_l}^n - 1 \Bigr) \; , 
\\
q_{j\rightarrow i}^{0^*, j} & = & 0 \; , 
\\
q_{j\rightarrow i}^{0^*, A_i} & \propto & 
\prod\limits_{l \in \partial j\backslash i} \Bigl( q_{l \rightarrow j}^{0, 0^*} + q_{l \rightarrow j}^{0^*, 0^*} \Bigr) \; .
\label{eq:qji0sAi}
\end{eqnarray}
In Eqs.~(\ref{eq:qji0Ai}) and (\ref{eq:qji0sAi}), the state of vertex $i$ is $A_i = k$ with $k$ being the index of one of its nearest neighbors except for $j$ (namely, $A_i \in \partial i \backslash j$). The same restriction applies to the following Eqs.~(\ref{eq:qji_iAi}) and (\ref{eq:qji_kAi}). Notice that $q_{j\rightarrow i}^{0, 0} = q_{j\rightarrow i}^{0, 0^*}$ and $q_{j\rightarrow i}^{0^*, 0} = q_{j\rightarrow i}^{0^*, 0^*}$.

If the state of vertex $j$ is taking $A_j=i$ in $q_{j\rightarrow i}^{A_j, A_i}$, we have
\begin{eqnarray}
q_{j\rightarrow i}^{i, 0} & = &  q_{j\rightarrow i}^{i, 0^*} = q_{j\rightarrow i}^{i, j}  =  0 \; , 
\\
q_{j\rightarrow i}^{i, A_i} & \propto & e^{-\beta}
\prod\limits_{k\in \partial j \backslash i} \sum\limits_{A_k} \Bigl[ \delta_{A_k}^0 q_{k\rightarrow j}^{0, i} + \delta_{A_k}^{0^*} q_{k\rightarrow j}^{0^*, i} + \delta_{A_k}^j q_{k\rightarrow j}^{j, i} + \sum\limits_{m\in \partial k\backslash  j} \delta_{A_k}^m q_{k\rightarrow j}^{m, i} \Bigr] 
\nonumber \\
& & \quad \times \; \Theta\Bigl( \sum\limits_{l\in \partial j\backslash i} (\delta_{A_l}^j + \delta_{A_l}^{0^*} ) - 1 \Bigr) \; .
\label{eq:qji_iAi}
\end{eqnarray}

If the state of vertex $j$ is taking $A_j=k \in \partial j \backslash i$ in $q_{j\rightarrow i}^{A_j, A_i}$, we have
\begin{eqnarray}
q_{j\rightarrow i}^{k, 0} & \propto & e^{-\beta} 
\Bigl[ \sum\limits_{m\in \partial k\backslash j} q_{k\rightarrow j}^{m, k} \Bigr]
\prod\limits_{l\in \partial j \backslash i, k} \sum\limits_{A_l} \Bigl[ \delta_{A_l}^0 q_{l\rightarrow j}^{0, k} + \delta_{A_l}^{0^*} q_{l\rightarrow j}^{0^*, k} + \delta_{A_l}^j q_{l\rightarrow j}^{j, k} + \sum\limits_{n\in \partial l\backslash  j} \delta_{A_l}^n q_{l\rightarrow j}^{n, k} \Bigr] 
\nonumber \\
& & \quad \quad \times \; \Theta\Bigl( \sum\limits_{p\in \partial j\backslash i,k}  (\delta_{A_p}^j + \delta_{A_p}^{0^*} ) - 1 \Bigr) \; ,
\\
q_{j\rightarrow i}^{k, 0^*} & \propto & e^{-\beta} 
\Bigl[ \sum\limits_{m\in \partial k\backslash j} q_{k\rightarrow j}^{m, k} \Bigr]
\prod\limits_{l\in \partial j \backslash i, k} \Bigl[ q_{l\rightarrow j}^{0, k} + q_{l\rightarrow j}^{0^*, k} + q_{l\rightarrow j}^{j, k} + \sum\limits_{n\in \partial l\backslash  j} q_{l\rightarrow j}^{n, k} \Bigr]  \; , 
\\
q_{j\rightarrow i}^{k, j} & \propto & e^{-\beta}  \Bigl[ \sum\limits_{m\in \partial k\backslash j} q_{k\rightarrow j}^{m, k} \Bigr] \prod\limits_{l\in \partial j \backslash i, k}  \Bigl[ q_{l\rightarrow j}^{0, k} + q_{l\rightarrow j}^{0^*, k} + q_{l\rightarrow j}^{j, k} + \sum\limits_{n\in \partial l\backslash  j} q_{l\rightarrow j}^{n, k} \Bigr] \; ,
\\
q_{j\rightarrow i}^{k, A_i} & \propto & e^{-\beta} 
\Bigl[ \sum\limits_{m\in \partial k\backslash j} q_{k\rightarrow j}^{m, k} \Bigr] \prod\limits_{l\in \partial j \backslash i, k} \sum\limits_{A_l} \Bigl[ \delta_{A_l}^0  q_{l\rightarrow j}^{0, k} + \delta_{A_l}^{0^*}  q_{l\rightarrow j}^{0^*, k} + \delta_{A_l}^j q_{l\rightarrow j}^{j, k} + \sum\limits_{n\in \partial l\backslash  j} \delta_{A_l}^n q_{l\rightarrow j}^{n, k} \Bigr] 
\nonumber \\
& & \quad \quad \times \; 
\Theta\Bigl( \sum\limits_{p\in \partial j\backslash i,k}  (\delta_{A_p}^j + \delta_{A_p}^{0^*} ) - 1 \Bigr) \; .
\label{eq:qji_kAi}
\end{eqnarray}
Notice that $q_{j\rightarrow i}^{k, 0^*} = q_{j\rightarrow i}^{k, j}$ and $q_{j\rightarrow i}^{k, 0} = q_{j\rightarrow i}^{k, A_i}$ for any $A_i \in \partial i\backslash j$.

The normalization constant for the cavity probabilities $q_{j\rightarrow i}^{A_j, A_i}$ is determined by
\begin{eqnarray}
  & &  q_{j\rightarrow i}^{0, 0} + q_{j\rightarrow i}^{0, 0^*} + q_{j\rightarrow i}^{0^*, 0} + q_{j\rightarrow i}^{0^*, 0^*} + \sum\limits_{k\in \partial i\backslash j} \bigl( q_{j\rightarrow i}^{0, k} \nonumber \\
     & & \quad + q_{j\rightarrow i}^{0^*, k} + q_{j\rightarrow i}^{i, k} \bigr) + \sum\limits_{l\in \partial j\backslash i} \bigl(
     q_{j\rightarrow i}^{l, 0} + q_{j\rightarrow i}^{l, 0^*} +   q_{j\rightarrow i}^{l, i} \bigr)
     +  \sum\limits_{l \in \partial j\backslash i} \sum\limits_{k \in \partial i\backslash j}  q_{j\rightarrow i}^{l, k}  = 1 \; .
\end{eqnarray}

The marginal probability $q_i^{A_i}$ quantifies the probability of observing vertex $i$ in state $A_i$ in an equilibrium configuration. It could be estimated according to
\begin{eqnarray}
q_i^0 & = & \frac{1}{z_i} \prod\limits_{j\in \partial i} \sum\limits_{A_j} \Bigl[ \delta_{A_j}^0 q_{j\rightarrow i}^{0, 0} + \delta_{A_j}^{0^*} q_{j\rightarrow i}^{0^*, 0} + \sum\limits_{k\in \partial j\backslash  i} \delta_{A_j}^k q_{j\rightarrow i}^{k, 0} \Bigr] \Theta\Bigl( \sum\limits_{l\in \partial i} \sum\limits_{m\in \partial l\backslash i} \delta_{A_l}^m - 2 \Bigr) \; , 
\\
q_i^{0^*} & = &  \frac{1}{z_i} \prod\limits_{j\in \partial i} \sum\limits_{A_j} \Bigl[ \delta_{A_j}^0 q_{j\rightarrow i}^{0, 0} + \delta_{A_j}^{0^*} q_{j\rightarrow i}^{0^*, 0} + \sum\limits_{k\in \partial j\backslash  i} \delta_{A_j}^k q_{j\rightarrow i}^{k, 0^{*}} \Bigr] I\Bigl( \sum\limits_{l\in \partial i} \sum\limits_{m\in \partial l\backslash i} \delta_{A_l}^m - 1 \Bigr) \; , 
\\
q_i^{j} & = & \frac{1}{z_i} e^{-\beta}  \Bigl[ \sum\limits_{k\in \partial j\backslash i} q_{j\rightarrow i}^{k, j} \Bigr] \prod\limits_{l\in \partial i \backslash j} \sum\limits_{A_l} \Bigl[ \delta_{A_l}^0  q_{l\rightarrow i}^{0, j} + \delta_{A_l}^{0^*}  q_{l\rightarrow i}^{0^*, j} + \delta_{A_l}^i q_{l\rightarrow i}^{i, j} + \sum\limits_{n\in \partial l\backslash  i} \delta_{A_l}^n q_{l\rightarrow i}^{n, j} \Bigr]  
\nonumber \\
& & \quad \quad \quad \quad \times 
\Theta\Bigl( \sum\limits_{p\in \partial i\backslash j}  \bigl( \delta_{A_p}^i + \delta_{A_p}^{0^*} \bigr) - 1 \Bigr) \; ,
\end{eqnarray}
in which the normalization constant $z_i$ is determined by
\begin{equation}
    q_i^0 + q_i^{0^*} + \sum\limits_{j\in \partial i} q_i^j = 1 \; .
\end{equation}

\subsection{Coarse-grained belief propagation equations}

Based on the BP equations of the previous subsection, we now define five coarse-grained cavity messages, $Q_{j\rightarrow i}^0$, $Q_{j\rightarrow i}^1$, $Q_{j\rightarrow i}^2$, $Q_{j\rightarrow i}^3$, and $Q_{j\rightarrow i}^4$, in the following way:
\begin{eqnarray}
Q_{j\rightarrow i}^{0} & \equiv &
\frac{1}{2} \Bigl( q_{j\rightarrow i}^{0, 0} + q_{j\rightarrow i}^{0^* , 0} + q_{j\rightarrow i}^{0, 0^*} + q_{j\rightarrow i}^{0^*, 0^*} \Bigr) \; , 
\label{eq:Qji0def}
\\
Q_{j\rightarrow i}^1 & \equiv &
\frac{1}{d_i-1} \sum\limits_{k\in \partial i\backslash j} q_{j\rightarrow i}^{0, k} 
\; , 
\label{eq:Qji1def}
\\
Q_{j\rightarrow i}^2 & \equiv &
\frac{1}{d_i - 1} \sum\limits_{k\in \partial i\backslash j}  \Bigl[ q_{j\rightarrow i}^{0^*, k} + q_{j\rightarrow i}^{i, k} \Bigr] 
\; , 
\label{eq:Qji2def}
\\
Q_{j\rightarrow i}^3 & \equiv &
\frac{1}{d_i} \sum\limits_{l\in \partial j\backslash i} \Bigl[ q_{j\rightarrow i}^{l, 0} +  \sum\limits_{k\in \partial i\backslash j} q_{j\rightarrow i}^{l, k} \Bigr] 
\; , 
\label{eq:Qji3def}
\\
Q_{j\rightarrow i}^4 & \equiv & \frac{1}{2} \sum\limits_{l\in \partial j\backslash i}  \Bigl[ q_{j\rightarrow i}^{l, 0^*} +  q_{j\rightarrow i}^{l, j} \Bigr]
\; .
\label{eq:Qji4def}
\end{eqnarray}

The values of these coarse-grained cavity probabilities could be self-consistently determined by a set of coarse-grained BP equations:
\begin{eqnarray}
Q_{j\rightarrow i}^{0} & = &
\frac{1}{z_{j\rightarrow i}} \biggl\{ \prod\limits_{l\in \partial j\backslash i} \sum\limits_{c_l} \Bigl( \delta_{c_l}^0 Q_{l\rightarrow j}^0 + \delta_{c_l}^3 Q_{l\rightarrow j}^3 \Bigr) \Theta\Bigl( \sum\limits_{m\in \partial j\backslash i} \delta_{c_m}^3 - 2 \Bigr)
\nonumber \\
& & \quad \quad \quad + \prod\limits_{l\in \partial j\backslash i} \sum\limits_{c_l} \Bigl( \delta_{c_l}^0 Q_{l\rightarrow j}^0 + \delta_{c_l}^4 Q_{l\rightarrow j}^4 \Bigr) I\Bigl( \sum\limits_{m\in \partial j\backslash i} \delta_{c_m}^4 - 1 \Bigr) \biggr\} \; ,
\label{eq:Qji0}
\\
Q_{j\rightarrow i}^1 & = &
\frac{1}{z_{j\rightarrow i}} \prod\limits_{l\in \partial j\backslash i} \sum\limits_{c_l} \Bigl( \delta_{c_l}^0 Q_{l\rightarrow j}^0 + \delta_{c_l}^3 Q_{l\rightarrow j}^3 \Bigr) \Theta\Bigl( \sum\limits_{m\in \partial j\backslash i} \delta_{c_m}^3 - 1 \Bigr) \; ,
\label{eq:Qji1}
\\
Q_{j\rightarrow i}^2 & = &
\frac{1}{z_{j\rightarrow i}} \biggl\{ 
 e^{-\beta}\prod\limits_{l\in \partial j\backslash i} \sum\limits_{c_l} \Bigl( \delta_{c_l}^1 Q_{l\rightarrow j}^1 + \delta_{c_l}^2 Q_{l\rightarrow j}^2 + \delta_{c_l}^3 Q_{l\rightarrow j}^3 \Bigr) \; \Theta\Bigl( \sum\limits_{m\in \partial j\backslash i} \delta_{c_m}^2 - 1 \Bigr) 
\nonumber \\
& & \quad \quad \quad + \; 
\prod\limits_{l\in \partial j\backslash i} Q_{l\rightarrow j}^0 
\biggr\} \; ,
\label{eq:Qji2}
\\
Q_{j\rightarrow i}^3 & = &
\frac{1}{z_{j\rightarrow i}} e^{-\beta} \sum\limits_{l\in \partial j\backslash i} Q_{l\rightarrow j}^4 \prod\limits_{m\in \partial j \backslash i, l} \sum\limits_{c_m} \Bigl( \delta_{c_m}^1 Q_{m \rightarrow j}^1 + \delta_{c_m}^2 Q_{m\rightarrow j}^2 + \delta_{c_m}^3 Q_{m\rightarrow j}^3\Bigr)
\nonumber \\
& & \quad \quad \quad \times \; 
\Theta\Bigl( \sum\limits_{n\in \partial j\backslash i, l} \delta_{c_n}^2  - 1 \Bigr)
\; ,
\label{eq:Qji3}
\\
Q_{j\rightarrow i}^4 & = & 
\frac{1}{z_{j\rightarrow i}} e^{-\beta} \sum\limits_{l \in \partial j\backslash i} Q_{l\rightarrow j}^4 \prod\limits_{m \in \partial j \backslash i, l} \Bigl( Q_{m\rightarrow j}^1 + Q_{m\rightarrow j}^2 + Q_{m\rightarrow j}^3 \Bigr)  
\; .
\label{eq:Qji4}
\end{eqnarray}
The normalization constant $z_{j\rightarrow i}$ is determined by the condition
\begin{equation}
 2 Q_{j\rightarrow i}^0 + (d_i-1) Q_{j\rightarrow i}^1 + (d_i -1) Q_{j\rightarrow i}^2 + d_i  Q_{j\rightarrow i}^3 + 2  Q_{j\rightarrow i}^4  = 1 \; .
 \label{eq:zjinormal}
\end{equation}

\subsection{Thermodynamic quantities}

The thermodynamic quantities such as the energy density $\rho$, free energy density $f$, and entropy density $s$ only depend on the coarse-grained cavity probabilities.  First, the marginal probability $Q_i^{0}$ of vertex $i$ not being a member of the backbone is
equal to $(q_i^0 + q_i^{0^*})$ and its expression is
\begin{eqnarray}
 Q_i^{0}  & = & \frac{1}{z_i} \biggl\{  \prod\limits_{j\in \partial i} \sum\limits_{c_j} \Bigl( \delta_{c_j}^0 Q_{j\rightarrow i}^0 + \delta_{c_j}^3 Q_{j\rightarrow i}^3 \Bigr) \; \Theta\Bigl( \sum\limits_{k\in \partial i} \delta_{c_k}^3 - 2 \Bigr)
 \nonumber \\
 & & \quad \quad + \prod\limits_{j\in \partial i} \sum\limits_{c_j} \Bigl( \delta_{c_j}^0 Q_{j\rightarrow i}^0 + \delta_{c_j}^4 Q_{j\rightarrow i}^4 \Bigr) \; I\Bigl( \sum\limits_{k\in \partial i} \delta_{c_k}^4 - 1 \Bigr) \biggr\} \; .
\end{eqnarray}
In the above expression, we denote by $c_j \in \{0,1,2,3,4\}$ the coarse-grained state associated with vertex $j$. Similarly, 
the marginal probability $Q_i^{b}$ of vertex $i$ being a member of the backbone is equal to $\sum_{j \in \partial i} q_i^j$ and its expression is
\begin{eqnarray}
 Q_i^{b}  & = & \frac{e^{-\beta}}{z_i} \sum\limits_{j\in \partial i} Q_{j\rightarrow i}^4 \prod\limits_{k\in \partial i\backslash j} \sum\limits_{c_k} \Bigl( \delta_{c_k}^1 Q_{k\rightarrow i}^1 + \delta_{c_k}^2 Q_{k\rightarrow i}^2 + \delta_{c_k}^3 Q_{k\rightarrow i}^3 \Bigr) \Theta\Bigl( \sum\limits_{l\in \partial i\backslash j} \delta_{c_l}^2 - 1 \Bigr) \; .
 \label{eq:Qjb}
\end{eqnarray}
The normalization constant $z_i$ ensures that $Q_i^0 + Q_i^{b} = 1$, and its expression is
\begin{eqnarray}
z_i & = & \prod\limits_{j\in \partial i} \sum\limits_{c_j} \Bigl( \delta_{c_j}^0 Q_{j\rightarrow i}^0 + \delta_{c_j}^3 Q_{j\rightarrow i}^3 \Bigr) \Theta\Bigl( \sum\limits_{k\in \partial i} \delta_{c_k}^3 - 2 \Bigr)
\nonumber \\
& & + \prod\limits_{j\in \partial i} \sum\limits_{c_j} \Bigl( \delta_{c_j}^0 Q_{j\rightarrow i}^0 + \delta_{c_j}^4 Q_{j\rightarrow i}^4 \Bigr) I\Bigl( \sum\limits_{k\in \partial i} \delta_{c_k}^4 - 1 \Bigr) 
\nonumber \\
& & + e^{-\beta} \sum\limits_{j\in \partial i} Q_{j\rightarrow i}^4 \prod\limits_{k\in \partial i\backslash j} \sum\limits_{c_k} \Bigl( \delta_{c_k}^1 Q_{k\rightarrow i}^1 + \delta_{c_k}^2 Q_{k\rightarrow i}^2 + \delta_{c_k}^3 Q_{k\rightarrow i}^3 \Bigr) \Theta\Bigl( \sum\limits_{l\in \partial i\backslash j} \delta_{c_l}^2 - 1 \Bigr) \; .
\end{eqnarray}
The energy density $\rho$ is simply the relative size of the backbone,
\begin{equation}
    \rho = \frac{1}{N} \sum\limits_{i=1}^{N} Q_i^{b} \; .
\end{equation}

The total free energy $F(\beta)$ of the system is evaluated through
\begin{equation}
F(\beta) = \sum_{i=1}^{N} f_{i} - \sum_{(i,j)\in G}f_{(i,j)} \; ,
\end{equation}
where $f_{i} = -(1/\beta) \ln (z_i)$ is the free energy contribution caused by the constraint associated with vertex $i$ and $f_{(i, j)}$ is the free energy contribution of an edge $(i, j)$. The expressions for these two free energy contributions are
\begin{eqnarray}
f_i & = & 
-\frac{1}{\beta} \ln\biggl\{\prod\limits_{j\in \partial i} \sum\limits_{c_j} \Bigl( \delta_{c_j}^0 Q_{j\rightarrow i}^0 + \delta_{c_j}^3 Q_{j\rightarrow i}^3 \Bigr) \Theta\Bigl( \sum\limits_{k\in \partial i} \delta_{c_k}^3 - 2 \Bigr) 
\nonumber \\
& & \quad + \prod\limits_{j\in \partial i} \sum\limits_{c_j} \Bigl( \delta_{c_j}^0 Q_{j\rightarrow i}^0 + \delta_{c_j}^4 Q_{j\rightarrow i}^4 \Bigr) I\Bigl( \sum\limits_{k\in \partial i} \delta_{c_k}^4 - 1 \Bigr)
\nonumber \\
& & \quad + e^{-\beta}\sum\limits_{j\in \partial i} Q_{j\rightarrow i}^4 \prod\limits_{k\in \partial i\backslash j} \sum\limits_{c_k} \Bigl( \delta_{c_k}^1 Q_{k\rightarrow i}^1 + \delta_{c_k}^2 Q_{k\rightarrow i}^2 + \delta_{c_k}^3 Q_{k\rightarrow i}^3 \Bigr) \Theta\Bigl( \sum\limits_{l\in \partial i\backslash j} \delta_{c_l}^2 - 1 \Bigr) \biggr\} \; ,
\nonumber \\
& &  
\end{eqnarray}
and
\begin{equation}
f_{(i,j)} = -\frac{1}{\beta} \ln\Bigl[Q_{j\rightarrow i}^{0}Q_{i\rightarrow j}^{0}+Q_{j\rightarrow i}^{1}Q_{i\rightarrow j}^{3}+Q_{j\rightarrow i}^{3}Q_{i\rightarrow j}^{1}+Q_{j\rightarrow i}^{3}Q_{i\rightarrow j}^{3}+Q_{j\rightarrow i}^{2}Q_{i\rightarrow j}^{4}+Q_{j\rightarrow i}^{4}Q_{i\rightarrow j}^{2}\Bigr] \; .
\end{equation}
%
%
The free energy density $f(\beta)$ is then evaluated through
\begin{equation}
    f(\beta) = \frac{1}{N} F(\beta) = \frac{1}{N} \sum\limits_{i} f_i - \frac{1}{N} \sum\limits_{(i, j)\in G} f_{(i, j)} \; .
\end{equation}
The entropy density $s$ is evaluated from the energy density $\rho$ and free energy density $f$ as
\begin{equation}
    s = \beta \rho - \beta f \; .
\end{equation}

\section{Brief description of the numerical simulation method}

\subsection{Belief propagation on a single graph instance}

To obtain a fixed-point solution of the coarse-grained BP equations (\ref{eq:Qji0})--(\ref{eq:Qji4}) on a single graph instance $G$, we adopt the simple and commonly used iteration method. The inverse temperature parameter is first fixed to a certain value $\beta$, and the cavity messages $Q_{j\rightarrow i}^{c_j}$ for $c_j \in \{0, \ldots, 4\}$ on all the edges are properly initialized. At each step of the BP iteration process, then we examine all the vertices $j$ of the graph in a random sequential order. For each vertex $j$, we update the output messages $Q_{j\rightarrow i}^{c_j}$ to all its nearest neighbors $j$ according to the coarse-grained BP equations, but with a damping factor $\eta$ to reduce oscillatory behavior. For example, the damped iteration of Eq.~(\ref{eq:Qji4}) reads
\begin{equation}
Q_{j\rightarrow i}^4 \; \Leftarrow \; (1- \eta) \; Q_{j\rightarrow i}^4 + \eta \;
\frac{1}{z_{j\rightarrow i}} e^{-\beta} \sum\limits_{l \in \partial j\backslash i} Q_{l\rightarrow j}^4 \prod\limits_{m \in \partial j \backslash i, l} \Bigl( Q_{m\rightarrow j}^1 + Q_{m\rightarrow j}^2 + Q_{m\rightarrow j}^3 \Bigr)  
\; .
\end{equation}
In our numerical simulations we choose a value of $\eta = 0.85$ for the damping factor. 

\subsection{Population dynamics for an ensemble of random graphs}

To compute ensemble-averaged thermodynamic quantities such as the mean energy density and free energy density, we perform population dynamics simulations for a given random graph ensemble characterized by the degree distribution $P(d_i)$. For a regular random graph ensemble this distribution is simply $P(d_i) = \delta_{d_i}^K$ (every vertex having the same degree $K$), while for the modified Erd\"os-R\'enyi graph ensemble this distribution is governed by Eq.~(\ref{eq:ERpd}).

An array $\mathcal{M}$ of length $L$ (say $L = 10^4$) is created in the computer memory. Each element of this array stores a $5$-dimensional cavity probability vector $(Q_{j\rightarrow i}^0, \ldots, Q_{j\rightarrow i}^4)$ and it is randomly initialized. Then at each elementary update step of the population dynamics: (1) a random degree $d_i$ is generated from the degree distribution $P(d_i)$ and this degree is assigned to a vertex $i$;  (2) a number $d_i$ of nearest neighbors $j$ are created for vertex $i$, and the cavity message vector $Q_{j\rightarrow i}$ of each vertex $j$ to $i$ being drawn uniformly at random from the array $\mathcal{M}$ with replacement; (3) a number $d_i$ of output cavity message vectors $Q_{i\rightarrow j}$ are computed following the coarse-grained BP equations (\ref{eq:Qji0})--(\ref{eq:Qji4}) for the nearest neighboring vertices $j$; (4) replace $d_i$ randomly chosen elements of the array $\mathcal{M}$ by these newly computed cavity probability vectors.

\section{Modified message-passing equations for the {\tt BBQ} algorithm}

Starting with a connected graph $G$ as input and an initially empty vertex set $\Gamma$, the {\tt BBQ} message-passing algorithm attempts to construct an approximately minimum connected dominating set and the associated backbone subgraph by sequentially adding vertices to $\Gamma$. Each vertex of set $\Gamma$ is referred to as an occupied (or backbone) vertex. During the running of the {\tt BBQ} process, the newly assigned backbone vertices are deleted from the graph $G$ together with the attached edges. In the following discussions, the degree $d_i$ of a vertex $i$ means its active degree, namely the number of nearest vertices of $i$ in the current graph $G$. Similarly the vertex set $\partial i$ means the set of nearest neighbors of vertex $i$ in the current graph $G$.

The coarse-grained BP iterations are performed on all the edges $(i, j)$ of the current graph $G$. The edges between the backbone vertices and those between the backbone and the current active graph $G$ are all ignored. The partially completed backbone is treated as a single effective occupied vertex, and it serves as the pointing target for those nearest-neighboring interface vertices in the current active graph $G$.

Consider a vertex $j$ in the current graph $G$. There are the following three distinct situations.

First, if this vertex $j$ is not staying at the interface between graph $G$ and the backbone subgraph (i.e., it is not connected to any vertices in the set $\Gamma$), then the output cavity message vectors $Q_{j\rightarrow i}$ to all its nearest neighbors $i$ are updated according to Eqs.~(\ref{eq:Qji0})--(\ref{eq:Qji4}).

Second, if one of the original nearest neighbors of vertex $j$ is now belonging to the vertex set $\Gamma$ and is missing in the current vertex set $\partial j$, then the coarse-grained BP equations for $Q_{j\rightarrow i}$ for any vertex $i \in \partial j$ will be slightly modified as
\begin{eqnarray}
Q_{j\rightarrow i}^{0} & = & 
\frac{1}{z_{j\rightarrow i}} \biggl\{ \prod\limits_{l\in \partial j\backslash i} \sum\limits_{c_l} \Bigl( \delta_{c_l}^0 Q_{l\rightarrow j}^0 + \delta_{c_l}^3 Q_{l\rightarrow j}^3 \Bigr) \Theta\Bigl( \sum\limits_{l\in \partial j\backslash i} \delta_{c_l}^3 - 1 \Bigr)
\; + \; \prod\limits_{l\in \partial j\backslash i} Q_{l\rightarrow j}^0  \biggr\} 
\nonumber \\
& = & 
\frac{1}{z_{j\rightarrow i}} \prod\limits_{l\in \partial j\backslash i} \Bigl( Q_{l\rightarrow j}^0 + Q_{l\rightarrow j}^3 \Bigr) 
\; ,
\label{eq:Qji0mod1}
\\
Q_{j\rightarrow i}^1 & = &
\frac{1}{z_{j\rightarrow i}} \prod\limits_{l\in \partial j\backslash i} \Bigl(  Q_{l\rightarrow j}^0 + Q_{l\rightarrow j}^3 \Bigr) 
\; , 
\label{eq:Qji1mod1}
\\
Q_{j\rightarrow i}^2 & = &  0 
\; , 
\label{eq:Qji2mod1}
\\
Q_{j\rightarrow i}^3 & = &
\frac{1}{z_{j\rightarrow i}} e^{-\beta} \prod\limits_{l\in \partial j \backslash i} \sum\limits_{c_l} \Bigl( \delta_{c_l}^1 Q_{l \rightarrow j}^1 + \delta_{c_l}^2 Q_{l\rightarrow j}^2 + \delta_{c_l}^3 Q_{l\rightarrow j}^3\Bigr)
\; \Theta\Bigl( \sum\limits_{l\in \partial j\backslash i} \delta_{c_l}^2  - 1 \Bigr)
\; ,
\label{eq:Qji3mod1}
\\
Q_{j\rightarrow i}^4 & = & 
\frac{1}{z_{j\rightarrow i}} e^{-\beta} \prod\limits_{l \in \partial j \backslash i} \Bigl( Q_{l\rightarrow j}^1 + Q_{l\rightarrow j}^2 + Q_{l\rightarrow j}^3 \Bigr)  
\; .
\label{eq:Qji4mod1}
\end{eqnarray}
The reason for writing down Eq.~(\ref{eq:Qji2mod1}) is that vertex $j$ shall point to the backbone set $\Gamma$, so $Q_{j\rightarrow i}^2 = 0$ according to the definition (\ref{eq:Qji2def}). The iterative expressions (\ref{eq:Qji3mod1}) and (\ref{eq:Qji4mod1}) are  also written down under the same consideration. According to Eq.~(\ref{eq:zjinormal}), the normalization constant $z_{j\rightarrow i}$ is determined by 
\begin{equation}
 2 Q_{j\rightarrow i}^0 + (d_i-1) Q_{j\rightarrow i}^1 + d_i  Q_{j\rightarrow i}^3 + 2  Q_{j\rightarrow i}^4  = 1 \; ,
 \label{eq:zjinormalmod}
\end{equation}
with $d_i$ being the vertex degree of vertex $i$ in the current active graph $G$ (not the original unperturbed one).  The marginal probability of such a vertex $j$ not being a member of the backbone ($Q_j^0$) and the probability of it being a member of the backbone ($Q_j^b$) are, respectively,
\begin{eqnarray}
 Q_j^{0}  & = & \frac{1}{z_j} \biggl\{  \prod\limits_{k\in \partial j} \sum\limits_{c_k} \Bigl( \delta_{c_k}^0 Q_{k\rightarrow j}^0 + \delta_{c_k}^3 Q_{k\rightarrow j}^3 \Bigr) \; \Theta\Bigl( \sum\limits_{k\in \partial j} \delta_{c_k}^3 - 1 \Bigr)
 \; + \; \prod\limits_{k\in \partial j} Q_{k\rightarrow j}^0  \biggr\} 
 \nonumber \\
 & = &  \frac{1}{z_j} \prod\limits_{k\in \partial j} \Bigl( Q_{k\rightarrow j}^0 + Q_{k\rightarrow j}^3 \Bigr)
\; ,
 \\
Q_j^{b}  & = &
\frac{e^{-\beta}}{z_j} \prod\limits_{k\in \partial  j} \sum\limits_{c_k} \Bigl( \delta_{c_k}^1 Q_{k\rightarrow i}^1 + \delta_{c_k}^2 Q_{k\rightarrow i}^2 + \delta_{c_k}^3 Q_{k\rightarrow i}^3 \Bigr) \Theta\Bigl( \sum\limits_{l\in \partial i\backslash j} \delta_{c_l}^2 - 1 \Bigr) \; .
\label{eq:Qjbmod1}
\end{eqnarray}
The normalization constant $z_i$ ensures that $Q_i^0 + Q_i^{b} = 1$.

If vertex $j$ has two or even more nearest neighbors belonging to the backbone set $\Gamma$, then 
\begin{eqnarray}
Q_{j\rightarrow i}^{0} & = & 
\frac{1}{z_{j\rightarrow i}} \prod\limits_{l\in \partial j\backslash i} \Bigl( Q_{l\rightarrow j}^0 + Q_{l\rightarrow j}^3 \Bigr)
\; , 
\label{eq:Qji0mod2}
\\
Q_{j\rightarrow i}^1 & = &
\frac{1}{z_{j\rightarrow i}} \prod\limits_{l\in \partial j\backslash i}  \Bigl(  Q_{l\rightarrow j}^0 +  Q_{l\rightarrow j}^3 \Bigr) 
\label{eq:Qji1mod2}
\\
Q_{j\rightarrow i}^2 & = &
0 \; ,
\label{eq:Qji2mod2}
\\
Q_{j\rightarrow i}^3 & = &
\frac{1}{z_{j\rightarrow i}} e^{-\beta} \prod\limits_{l\in \partial j \backslash i} \sum\limits_{c_l} \Bigl( \delta_{c_l}^1 Q_{l \rightarrow j}^1 + \delta_{c_l}^2 Q_{l\rightarrow j}^2 + \delta_{c_l}^3 Q_{l\rightarrow j}^3\Bigr)
\; \Theta\Bigl( \sum\limits_{l\in \partial j\backslash i} \delta_{c_l}^2  - 1 \Bigr)
\; ,
\label{eq:Qji3mod2}
\\
Q_{j\rightarrow i}^4 & = & 
\frac{1}{z_{j\rightarrow i}} e^{-\beta} \prod\limits_{l \in \partial j \backslash i} \Bigl( Q_{l\rightarrow j}^1 + Q_{l\rightarrow j}^2 + Q_{l\rightarrow j}^3 \Bigr)  
\; .
\label{eq:Qji4mod2}
\end{eqnarray}
The normalization condition is the same as Eq.~(\ref{eq:zjinormalmod}). The marginal probability of such a vertex $j$ of not being a member of the backbone ($Q_j^0$) and the probability of it being a member of the backbone ($Q_j^b$) are, respectively,
\begin{eqnarray}
Q_j^{0}  & = & \frac{1}{z_j}  \prod\limits_{i\in \partial j} \Bigl( Q_{i\rightarrow j}^0 + Q_{i\rightarrow j}^3 \Bigr) \;,
\\
Q_{j}^{b} & = & 
\frac{e^{-\beta}}{z_j} \prod\limits_{k\in \partial  j} \sum\limits_{c_k} \Bigl( \delta_{c_k}^1 Q_{k\rightarrow i}^1 + \delta_{c_k}^2 Q_{k\rightarrow i}^2 + \delta_{c_k}^3 Q_{k\rightarrow i}^3 \Bigr) \Theta\Bigl( \sum\limits_{l\in \partial i\backslash j} \delta_{c_l}^2 - 1 \Bigr) \; .
\label{eq:Qjbmod2}
\end{eqnarray}
The normalization constant $z_j$ is again determined by $Q_j^0 + Q_j^{b} = 1$.

It is interesting to notice that the message-passing equations (\ref{eq:Qji0mod2})--(\ref{eq:Qjbmod2}) are identical to the corresponding equations (\ref{eq:Qji0mod1})--(\ref{eq:Qjbmod1}). So actually we do not need to distinguish whether the vertex $j$ has only one nearest neighbors in the backbone or it has many such nearest neighbors. This could further simplify the message-passing algorithm.

\bibliography{reflist}

\end{document}